\documentclass[aps,prc,twocolumn,showpacs,preprintnumbers,amsmath,amssymb,superscriptaddress,nofootinbib]{revtex4-2}
\usepackage{graphicx} 
\usepackage[dvipdfmx]{epsfig} 
\usepackage{bm}
\usepackage{ulem}
\usepackage{amsmath,color}

\begin{document}
\title{Glauber-theory analysis of nuclear reactions on $^{12}$C target with variational Monte Carlo wave functions}
\author{W. Horiuchi}
\email{whoriuchi@omu.ac.jp}
\affiliation{Department of Physics, Osaka Metropolitan University, Osaka 558-8585, Japan}
\affiliation{Nambu Yoichiro Institute of Theoretical and Experimental Physics (NITEP), Osaka Metropolitan University, Osaka 558-8585, Japan}
\affiliation{RIKEN Nishina Center, Wako 351-0198, Japan}

\author{Y. Suzuki}
\affiliation{Department of Physics, Niigata University, Niigata 950-2181, Japan}
\affiliation{RIKEN Nishina Center, Wako 351-0198, Japan}

\author{R.~B. Wiringa}
\affiliation{Physics Division, Argonne National Laboratory, Argonne, Illinois 60439, USA}

\begin{abstract}
The application of Glauber theory has been playing an increasingly 
important role with the study of unstable or exotic nuclei. Its adaptation to medium and high-energy 
nucleus-nucleus collisions  is severely limited because one has to evaluate the matrix 
elements of multiple-scattering operators.  The extraction of physical observables has been 
done using `approximate' Glauber theory whose validity is hard to evaluate. We  perform 
a full calculation of the matrix elements using Monte Carlo integration and analyze the elastic 
differential cross sections and the total reaction cross sections for $p+^{12}$C, $^{4,6}$He$+^{12}$C, and $^{12}$C$+^{12}$C collisions.  We use the variational Monte Carlo wave functions for $^{4,6}$He and $^{12}$C obtained by using realistic two- and three-nucleon potentials. 
We demonstrate the performance of the Glauber-theory calculations by 
comparing with available experimental data. We further  discuss the accuracy of
the conventional approximate methods in the light of the cumulant expansion for Glauber's phase-shift 
function. 
\end{abstract}
\maketitle

\section{Introduction}

The study of  unstable nuclei has been performed mostly by 
measuring the cross sections of radioactive nuclear reactions
at high energies~\cite{Tanihata85,Tanihata13,Tanaka24}.
Since the observed cross sections can be related to the wave
functions of those nuclei through Glauber theory~\cite{Glauber},  
one can obtain information on the structure of the unstable nuclei. 
Among the cross sections, the elastic scattering cross sections and the total 
reaction cross sections can unambiguously be related to the ground-state  wave functions of both 
the projectile and target nuclei. 

In Glauber theory the nucleus-nucleus elastic scattering
amplitude is obtained by integrating the profile 
function  over the impact parameter. The profile function 
is the matrix element of the multiple-scattering operator
between the product of the wave functions of the target
and projectile nuclei. The difficulty of its evaluation is due to 
the fact that the multiple-scattering operator
is  a product of pairwise nucleon-nucleon
scattering operators, resulting in an $A$-body operator, where $A$ is the sum of the mass numbers of the projectile and target nuclei. To avoid the calculation of the matrix
element of the $A$-body operator  several simplifying approximations
have been introduced over the years.
The problem of these approximations is that it is very difficult
to judge how good they are. 

The power of a Monte Carlo integration  (MCI) to 
evaluate the profile function was shown for the first time 
in Ref.~\cite{Varga02}, where  a microscopic three-$\alpha$ cluster 
model wave function was used for $^{12}$C. 
The purpose of the present paper is to focus on 
the total reaction and elastic scattering cross sections of 
$p+^{12}$C, $^{4}$He$+^{12}$C, 
$^{6}$He$+^{12}$C, and $^{12}$C$+^{12}$C using the variational Monte Carlo (VMC) wave function for $^{12}$C that is now available and has been used in studies of ($p$,$pN$) reactions \cite{Crespo20,Cravo24}, electron scattering from $^{12}$C \cite{Andreoli24}, and model studies of neutrinoless double-beta decay \cite{Weiss22}. 
Since the wave functions of the projectile and target nuclei are all described by the VMC method,  the present results can directly be compared to experiment unambiguously. 

The multiple-scattering operator contains both the nuclear potential and Coulomb potential terms. 
The latter term primarily contributes to the projectile-target Coulomb potential leading to the Rutherford scattering. However, there arises such a term that can contribute to the breakup 
of the projectile and target nuclei, the so-called Coulomb breakup effect. Though several attempts have been 
proposed regarding how to separate the Coulomb contributions into the two, no convincing recipe has 
been available for the composite-particle scatterings. We propose a physically 
plausible separation of the Coulomb contributions into the two terms.

As mentioned above, the profile function can be calculated with the MCI without recourse to any truncation of the $A$-body operators.  This advantage can be used to evaluate some 
approximations to the profile function including the optical-limit approximation (OLA)~\cite{Glauber,Suzuki03} and the nucleon-target formalism in the Glauber theory (NTG)~\cite{NTG} approximation. In addition we show that the cumulant expansion~\cite{Glauber, Kubo62, Hufner81, Ogawa92} of the profile function is a very 
powerful way to establish how fast the multiple-scattering operators converge.  We also show the 
relationship between the cumulants  and the $n$-particle densities of the projectile and 
target nuclei. 

The paper is organized as follows. 
Section~\ref{formalism.sec} summarizes the formulation of
the present work in three subsections: 
the fundamentals of Glauber theory (Sect.~\ref{Glauber.sec}),  
 details of the phase-shift function that is 
a crucial vehicle in Glauber theory (Sect.~\ref{phaseshift.fn}), and  
the formulas to calculate  elastic differential  cross sections
and total reaction cross sections 
(Sect.~\ref{cs.sec}). 
Section~\ref{VMC.sec}  explains the VMC
wave functions used in this paper.
Section~\ref{numerical.sec} briefly introduces  the  MCI to calculate 
the phase-shift function. 
Section~\ref{results.sec} presents our numerical results: 
the kinematics for  velocity and  momentum (Sect.~\ref{kinematics.sec}), 
details of the Coulomb breakup effects (Sect.~\ref{cbu.sec}), and the 
comparison of elastic differential cross sections 
and total reaction cross sections with experiment (Sect.~\ref{elastic-reaction.sec}). The last 
subsection is further divided into four parts: 
 $p+^{12}$C scatterings (Sect.~\ref{resultsC-p.sec}) , $^{12}$C$+^{12}$C scatterings  
(Sect.~\ref{resultsC-C.sec}), $^{4}$He$+^{12}$C scatterings (Sect.~\ref{results4He-C.sec}), 
and $^{6}$He$+^{12}$C scatterings (Sect.~\ref{results6He-C.sec}). 
Section~\ref{approx.sec} discusses some approximate ways to compute the profile function and 
evaluates the accuracy of those approximations in the light of the cumulant expansion. 
Section~\ref{conclusion.sec} gives a summary and outlook of the present work.

\section{Formalism}
\label{formalism.sec}

\subsection{Glauber theory for nucleus-nucleus scattering}
\label{Glauber.sec}

We start by reviewing  the basics of Glauber theory for high-energy nuclear collisions~\cite{Glauber,Yabana92,Suzuki03}. 
Let $H_P$ and $H_T$ denote  respectively  
the   Hamiltonians of the projectile and  target nuclei with  mass numbers, $A_P$ and $A_T$.  Their ground-state wave functions,  $\Psi_0^P$ and $\Psi_0^T$, satisfy 
\begin{align}
H_P\Psi_0^P(\{\bm{r}^P\})&=E_0^P\Psi_0^P(\{\bm{r}^P\}),
\notag \\
H_T\Psi_0^T(\{\bm{r}^T\})&=E_0^T\Psi_0^T(\{\bm{r}^T\}).
\end{align}
Here, e.g. $\{\bm{r}^P\}$ denotes a set of $\bm{r}_1^P, ..., \bm{r}_{A_P}^P$, where $\bm{r}_i^{P }$  is 
the $i$th single-nucleon coordinate of the projectile nucleus. Not all of them are independent as 
they satisfy $\sum_{i=1}^{A_P} \bm{r}_i^P=0$.   
The spin and isospin coordinates are not explicitly written for simplicity,  and the integration with respect to those coordinates is implicitly understood. The total Hamiltonian of the $P$+$T$ colliding system reads 
\begin{align}
  H=\frac{\bm{P}^2}{2M_{PT}}+H_P+H_T+\sum_{i=1}^{A_P}\sum_{j=1}^{A_T}
  V(\bm{R}+\bm{r}_i^P-\bm{r}_j^T),
\label{Hamil.eq}
\end{align}
where $\bm{P}$ and $\bm{R}$ are respectively the relative momentum and coordinate of the two nuclei 
and  $M_{PT}=(A_PA_T/(A_P+A_T))m_N$ is their reduced mass, where $m_N$ is the nucleon mass. The nucleon-nucleon ($NN$) interaction $V$ includes both  nuclear and 
Coulomb potentials, $V_{NN}$ and $V_C$.
  The three-nucleon forces acting between $P$ and $T$
  are ignored in Eq.~(\ref{Hamil.eq}) as they are expected
  to make small contributions. See the arguments on the three-nucleon forces in Sect.~\ref{VMC.sec}.

Let $\bm{R}$ be written as $\bm{R}=\bm{b}+Z\hat{\bm{z}}$ with the impact parameter vector $\bm{b}$ and  the unit vector $\hat{\bm{z}}$ in the beam direction.  {At $Z\to -\infty$ the relative wave function should be a pure incoming wave. To remove the rapid oscillation of the incident wave, we set the scattering wave function $\Psi$ to 
\begin{align}
  \Psi(\bm{R},\{\bm{r}^P\},\{\bm{r}^T\}) \equiv  e^{iKZ}\hat{\Psi}(\bm{R},\{\bm{r}^P\},\{\bm{r}^T\}),
  \label{eikonal.eq}
\end{align}
where $\hbar K$ is the initial momentum of the relative motion and $\hat{\Psi}$ satisfies the 
initial condition 
\begin{align}
& \hat{\Psi}(\bm{R},\{\bm{r}^P\},\{\bm{r}^T\})
 \to \Psi_0^P(\{\bm{r}^P\}) \Psi_0^T(\{\bm{r}^T\})
\label{initial.cond}
\end{align}
for $  Z \to -\infty$.
The equation of motion for $\hat{\Psi}$ reads 
\begin{widetext}
\begin{align}
  \left[vP_Z+\frac{\bm{P}^2}{2M_{PT}}+(H_P-E_0^P)+(H_T-E_0^T)+
  \sum_{i\in P}^{A_P}\sum_{j\in T}^{A_T}
  V(\bm{R}+\bm{r}_i^P-\bm{r}_j^T)\right]
  \hat{\Psi}(\bm{R},\{\bm{r}^P\},\{\bm{r}^T\})=0,
  \label{eikeq.eq}
\end{align}  
\end{widetext}
where $P_Z=\frac{\hbar}{i}\frac{\partial}{\partial Z}$ and $v=\hbar K/M_{PT}$ is the relative velocity. 

$\hat{\Psi}$ is expected to be a slowly varying function with respect to $\bm{R}$,
and one can omit $\bm{P}^2/(2M_{PT})$ 
as it is expected to be much smaller than $vP_Z$ (the eikonal approximation).
This assumption is valid when the fluctuation of the momentum is much smaller
than $\hbar K$.
As the fluctuation is related to the size of the system,
the condition that the eikonal approximation is valid can be set as
$Ka\gg 1$ with $a$ being the range of the potential
of the two colliding nuclei. 
The condition is usually fulfilled
when we consider the short-ranged nuclear force
at the incident energies of more than several tens of MeV.
Note, however,  that the Coulomb potential does not satisfy the condition
and needs a  treatment differently from the nuclear force, as will be discussed later. 
The eikonal approximation also implies the condition
of the low momentum transfer, $q=|\bm{q}| \ll K$. Here, $\bm{q}=\bm{K}^{\prime}-\bm{K}$ and   
$\hbar K^{\prime}$ is the final momentum of the relative motion. 

Equation~(\ref{eikeq.eq}) is further simplified by assuming
the adiabatic approximation that ignores  
the intrinsic Hamiltonians, $H_P-E_0^P$ and $H_T-E_0^T$, compared to $vP_Z$. They measure the excitation energies of the respective nuclei.  In their ground states the maximum kinetic energy of the nucleon is 
estimated to be two times the Fermi energy. That is, the adiabatic approximation assumes that the incident energy per nucleon is high enough compared to $80$~MeV. For a proton-nucleus case the proton energy must be much higher than 40~MeV. 
Once $\bm{P}^2/(2M_{PT})$, $H_P-E_0^P$, and $H_T-E_0^T$ are all omitted,   Eq.~(\ref{eikeq.eq}) reduces to a first-order differential equation for $\hat{\Psi}$. The solution satisfying the initial condition~(\ref{initial.cond}) is given by 
\begin{widetext}
\begin{align}
  \hat{\Psi}(\bm{R},\{\bm{r}^P\},\{\bm{r}^T\})
=\exp\left[\frac{1}{i\hbar v} \int_{-\infty}^Z dZ^\prime
\sum_{i=1}^{A_P}\sum_{j=1}^{A_T}
V(\bm{b}+Z^{\prime}\bm{\hat{\bm z}}+\bm{r}_i^P-\bm{r}_j^T)\right]\Psi_0^P(\{\bm{r}^P\})
  \Psi_0^T(\{\bm{r}^T\}).
\label{psi.hat}
\end{align}  
\end{widetext}
Substitution of Eq.~(\ref{psi.hat}) into Eq.~(\ref{eikonal.eq}) gives the desired scattering wave function. Note that it has no correct 
outgoing asymptotic form but should be regarded valid in the region where 
the projectile and target potential does not vanish.

The elastic scattering amplitude reads as~\cite{Glauber,Suzuki03}
\begin{align}
  F(q)=\frac{iK}{2\pi}\int d\bm{b}\,e^{-i\bm{q}\cdot\bm{b}} \,\Gamma_G(\bm{b}),
\label{Glamp.eq}
\end{align}
where $\Gamma_G(\bm{b})$ is the profile function
\begin{align}
  \Gamma_G(\bm{b})=1-e^{i\chi_G(\bm{b})}.
\label{profile.fn}
\end{align}
Here,  $\chi_G(\bm{b})$ is Glauber's phase-shift function (psf) defined by 
\begin{align}
 & e^{i\chi_G(\bm{b})}\nonumber \\
 & =\left<\Psi_0^P\Psi_0^T\right|
  \prod_{j=1}^{A_P}\prod_{k=1}^{A_T}
  e^{i\chi_{NN}(\bm{b}_{jk})+i\epsilon_j\epsilon_k\chi_{C}(\bm{b}_{jk})}
  \left|\Psi_0^P\Psi_0^T\right>\notag \\
&\equiv \left<  \prod_{j=1}^{A_P}\prod_{k=1}^{A_T}
  e^{i\chi_{NN}(\bm{b}_{jk})+i\epsilon_j\epsilon_k\chi_{C}(\bm{b}_{jk})}  \right>.
\label{sigma_r.eq}
\end{align}
By expressing each nucleon coordinate by $\bm{r}_j=\bm{s}_j+ z_j\hat{\bm{z}} $, $\bm{b}_{jk}$ is defined 
 by $\bm{b}+\bm{s}_j^P-\bm{s}_k^T$.  The label $\epsilon_j$  distinguishes either proton ($\epsilon_j=1$) or neutron ($\epsilon_j=0$).  
  The nuclear and Coulomb phases, $\chi_{NN}$ and $\chi_{C}$,  are respectively related to the  nuclear and Coulomb potentials by
\begin{align}
&\chi_{NN}(\bm{b})=-\frac{1}{\hbar v} 
\int_{-\infty}^{\infty} dz V_{NN} (\bm{b}+z \hat{\bm{z}}), 
\label{chi_NN} 
\\
&\chi_{C}(\bm{b})=-\frac{1}{\hbar v} \int_{-\infty}^{\infty} dz
  V_{C}(\bm{b}+z \hat{\bm{z}}).
\label{chi_pp}
\end{align}
The integral of Eq.~(\ref{sigma_r.eq}) extends over all the independent coordinates including the spin-isospin coordinates. That is, one has to carry out the multi-dimensional integration to obtain the psf. 

\subsection{Phase-shift function}
\label{phaseshift.fn}

\subsubsection{Nuclear phase}

$\chi_{NN}(\bm{b})$ or the $NN$ profile function
\begin{align}
  \Gamma_{NN}(\bm{b})=1- e^{i\chi_{NN}(\bm{b})}
\end{align}
is a key quantity to proceed further. It is related to the  $NN$ potential $V_{NN}$ as shown in Eq.~(\ref{chi_NN}). Though  progress has constantly been pursued for a unified  description of both structure and reactions starting from realistic $NN$ potentials (see,  e.g., Refs.~\cite{Vorabbi22,Baker22} for recent developments), it is too difficult to derive the profile function from realistic $V_{NN}$ because one needs it at high incident energies beyond  hadron  production thresholds. Rather one parametrizes $\Gamma_{NN}(\bm{b})$ consistently with the $NN$ scattering data as follows~\cite{Ray79}:
\begin{align}
  \Gamma_{NN}(\bm{b})=\frac{1-i\alpha_{NN}}{4\pi\beta_{NN}}\sigma^{\rm tot}_{NN}
  \exp\left(-\frac{b^2}{2\beta_{NN}}\right).
\label{profile.eq}
\end{align}
Because $\sigma^{\rm tot}_{NN}$, $\alpha_{NN}$, and $\beta_{NN}$
are respectively related to the  $NN$ total cross section,
the ratio of the real and imaginary parts
of the $NN$ scattering amplitude, and the angular distribution
of the $NN$ elastic scattering, they  are determined consistently with 
the $NN$ scattering data as a function of the incident energy~\cite{Horiuchi07, Ibrahim08}. 
We use the parameter set~\cite{Ibrahim08} that takes into account 
the difference between  $pn$ and $pp$ scattering properties. 
Note that $\Gamma_{pp}=\Gamma_{nn}$ and $\Gamma_{pn}=\Gamma_{np}$ are assumed.
The parameter set has been used in many examples
of proton-nucleus and nucleus-nucleus scatterings including unstable nuclei~\cite{Ibrahim08, Ibrahim09, Horiuchi10, Kaki12, Horiuchi12, Horiuchi14, Horiuchi15, Horiuchi16, Horiuchi17, Nagahisa18, Hatakeyama18, Hatakeyama19, Horiuchi20, Horiuchi20b, Choudhary20, Horiuchi21a, Horiuchi21b, Choudhary21, Horiuchi22a, Horiuchi22b, Makiguchi22, Horiuchi22c, Horiuchi23a, Horiuchi23b, Yamaguchi23, Singh24,Okada24, Barman25, Inakura25}.
         
The total nuclear term of the multiple-scattering operator (actually function) in Eq.~(\ref{sigma_r.eq}) reads
\begin{align}
  e^{i\chi_{N}^{\rm tot}(\bm{b},\{\bm{s}^P\},\{\bm{s}^T \})}\equiv
  \prod_{j=1}^{A_P}\prod_{k=1}^{A_T}
  \left[1-\Gamma_{NN}(\bm{b}_{jk})\right].
\label{nucl.part}
\end{align}
Here, e.g., $\{\bm{s}^P\}$ denotes a set of $\bm{s}_1^P, ..., \bm{s}_{A_P}^P$.

\subsubsection{Coulomb phase}

Here we discuss the total Coulomb potential term defined by
\begin{align}
  e^{i\chi_C^{\rm tot}(\bm{b},\{\bm{s}^P\},\{\bm{s}^T \})}\equiv  \prod_{j=1}^{A_P}\prod_{k=1}^{A_T}
  e^{i\epsilon_j\epsilon_k \chi_C(\bm{b}_{jk})}.
\end{align}
Since the Coulomb interaction is long-ranged, special care is needed for its implementation in Eq.~(\ref{sigma_r.eq}).
The phase of the Coulomb potential $V_C(r)=\frac{e^2}{r}$
is divergent (see Eq.~(\ref{chi_pp}))
\begin{align}
  \chi_C(b)=-\eta\int_{-\infty}^{\infty} dz\frac{1}{\sqrt{b^2+z^2}},
\end{align}
where $\eta=\frac{e^2}{\hbar v}$. 
The divergence is avoided  by  confining the Coulomb potential  
by the Heaviside step function $\Theta$,
\begin{align}
  V_C(r)=\frac{e^2}{r}\Theta(D-r),
\end{align}
where $D$ is a cut-off parameter. This leads to 
\begin{align}
  \chi_C(b)=-2\eta\Theta(D-b)
  \ln\frac{D+\sqrt{D^2-b^2}}{b}.
\end{align}
For  sufficiently large $D$, we obtain the usual expression
\begin{align}
  \chi_C(b)=2\eta\ln\frac{b}{2D}.
\end{align}
 The Coulomb phase due to the point-Coulomb potential responsible for the Rutherford scattering reads 
\begin{align}
  \chi_C^{\rm point}(b)=2\eta_{PT}\ln\frac{b}{2D},
\end{align}
where $\eta_{PT}=Z_PZ_T\eta$.
Microscopically the total Coulomb phase is expressed by 
\begin{align}
  \chi_C^{\rm tot} (\bm{b},\{\bm{s}^P\},\{\bm{s}^T \})=2\eta\sum_{j=1}^{A_P}\sum_{k=1}^{A_T}
 \epsilon_j\epsilon_k\ln\frac{|\bm{b}_{jk}|}{2D},
\end{align}
which  is decomposed to 
\begin{align}
  &\chi_C^{\rm tot}(\bm{b},\{\bm{s}^P\},\{\bm{s}^T \})=\chi_C^{\rm point}(b)+\Delta\chi_C(\bm{b},\{\bm{s}^P\},\{\bm{s}^T \}),\notag \\
  &\Delta\chi_C(\bm{b},\{\bm{s}^P\},\{\bm{s}^T \})=2\eta\sum_{j=1}^{A_P}\sum_{k=1}^{A_T}
\epsilon_j\epsilon_k  \ln\frac{|\bm{b}_{jk}|}{b}.
\label{Delta.chi_c}
\end{align}
$\Delta\chi_C(\bm{b},\{\bm{s}^P\},\{\bm{s}^T \})$ accounts for the deviation
from the Coulomb phase responsible for the Rutherford scattering.  Note that it  
is independent of the cut-off parameter $D$. 
Though we are tempted to set $e^{i\Delta\chi_C(\bm{b},\{\bm{s}^P\},\{\bm{s}^T \})}$ to the Coulomb breakup term of the multiple-scattering operator in Eq.~(\ref{sigma_r.eq}), it is not necessarily  
correct because then the cross section diverges   
due to the adiabatic approximation~\cite{NTG}. 
We assume that $\Delta\chi_C(\bm{b},\{\bm{s}^P\},\{\bm{s}^T \})$ takes the form of Eq.~(\ref{Delta.chi_c}) in so far as the charge distributions of the projectile and target nuclei  
overlap, otherwise it vanishes. That is, the Coulomb breakup term (CBU)  in Eq.~(\ref{sigma_r.eq}) is  set to
\begin{align}
&e^{i\Delta\chi_C(\bm{b},\{\bm{s}^P\},\{\bm{s}^T \})}\notag \\
&\   \equiv \prod_{j=1}^{A_P}\prod_{k=1}^{A_T}
 \left(\frac{|\bm{b}_{jk}|}{b}\right)^{2i\eta \epsilon_j\epsilon_k}\Theta(b_C-b)+\Theta(b-b_C).
\label{coulomb.part}
\end{align}
Here, $b_C$ is a cut-off impact parameter chosen to be a sum of the proton radii  of the projectile and target nuclei, $\sqrt{\frac{5}{3}}(r_p^P+r_p^T)$, where, e.g. $r_p^P$ is the root-mean-square (rms) proton radius of the projectile nucleus.

Using the relation
\begin{align}
&e^{i\chi_{N}^{\rm tot}(\bm{b},\{\bm{s}^P\},\{\bm{s}^T \})+i\chi_{C}^{\rm point}(\bm{b})+i\Delta\chi_C(\bm{b},\{\bm{s}^P\},\{\bm{s}^T \})}\notag \\
&= e^{i\chi_{C}^{\rm point}(\bm{b})} -e^{i\chi_{C}^{\rm point}(\bm{b})}
(1-e^{i\chi(\bm{b},\{\bm{s}^P\},\{\bm{s}^T \})}),
  \label{subC.eq}
\end{align}
where 
\begin{align}
&\chi(\bm{b},\{\bm{s}^P\},\{\bm{s}^T \})\notag \\
&=\chi_{N}^{\rm tot}(\bm{b},\{\bm{s}^P\},\{\bm{s}^T \})+
\Delta\chi_C(\bm{b},\{\bm{s}^P\},\{\bm{s}^T \}), 
\end{align}
we separate $\chi_C^{\rm point}(b)$ from the psf of Eq.~(\ref{sigma_r.eq}) as follows:   
\begin{align}
  e^{i\chi_G(\bm{b})}&=\left<
  e^{i\chi_{N}^{\rm tot}(\bm{b})+i\chi_{C}^{\rm point}(\bm{b})+i\Delta\chi_C(\bm{b})}
  \right>\notag \\
&=e^{i\chi_{C}^{\rm point}(\bm{b})}-e^{i\chi_{C}^{\rm point}(\bm{b})}(1-e^{i\chi(\bm{b})}),
\end{align}
where  $\chi(\bm{b})$ is defined by 
\begin{align}
e^{i\chi(\bm{b})}&=\left<e^{i\chi(\bm{b},\{\bm{s}^P\},\{\bm{s}^T \})}\right>\notag \\
&=\left<e^{i\chi_N^{\rm tot}(\bm{b},\{\bm{s}^P\},\{\bm{s}^T \})} e^{i\Delta\chi_C(\bm{b},\{\bm{s}^P\},\{\bm{s}^T \})}\right>,
\label{psf.final}
\end{align}
and  contains both nuclear  and  CBU phases. 

\subsection{Cross sections}
\label{cs.sec}

The elastic scattering amplitude of Eq.~(\ref{Glamp.eq}) is cast to 
\begin{widetext}
\begin{align}
  F(q)
  &=e^{-2i\eta_{PT}\ln (2KD)}
  \left\{F_C(q)+\frac{iK}{2\pi}\int d\bm{b}\,
  e^{-i\bm{q}\cdot\bm{b}+2i\eta_{PT}\ln(Kb)}
  (1-e^{i\chi(\bm{b})})
  \right\},
\label{elastic.X}
\end{align}
\end{widetext}
where $F_C(q)$ is the Rutherford scattering amplitude 
\begin{align}
  F_C(q)=-\frac{2K\eta_{PT}}{q^2}e^{-2i\eta_{PT}
    \ln(\frac{q}{2K})+2i\,{\rm arg}\,\Gamma(1+i\eta_{PT})}.  
\end{align} 
The elastic differential  cross section  reads as 
\begin{align}
  \frac{d\sigma}{d\Omega}=|F(q)|^2,
\end{align}
and the total reaction cross section  is given by integrating  the absorption 
probability as ~\cite{Glauber,Yabana92,Suzuki03}
\begin{align}
\sigma_R=\int d\bm{b}\,\left(1-|e^{i\chi(\bm{b})}|^2\right).
\label{reaction.X}
\end{align}
Note that both cross sections are independent of the cut-off parameter $D$, and that 
the integration with respect to $\bm{b}$ gives no divergence.

The  value of $\eta_{PT}$ can  be as large as $\approx 1$
even for the light-ion collisions to be discussed here 
when the incident energy $E$ is low, e.g.,
$\frac{v}{c}\approx 0.3$ for $E\approx 50$~MeV/nucleon.
The trajectory of the projectile nucleus may be bent in that case, which increases 
the elastic scattering cross sections at backward angles and decreases 
the total reaction cross section.
The trajectory bend is taken into account
by replacing the impact parameter $b$ in  $\chi(\bm{b})$ with
the distance of the closest approach in the classical Coulomb trajectory~\cite{Vitturi87}
\begin{align}
  Kb \to \sqrt{(Kb)^2+\eta_{PT}^{2}}+\eta_{PT}.
\end{align}
This correction will be referred to as the Coulomb trajectory correction (CTC) hereafter.

\section{Variational Monte Carlo wave functions}
\label{VMC.sec}

In this work, we employ the wave function of $^{12}$C and $^{4,6}$He,
which are taken from VMC calculations for a Hamiltonian consisting of
nonrelativistic nucleon kinetic energy, the Argonne $v_{18}$ two-nucleon 
potential~\cite{WSS95}, and the Urbana X three-nucleon potential~\cite{WSPC14}:
\begin{align}
H = { \sum_{i} K_i } + { { \sum_{i<j}} v_{ij} }
+ { \sum_{i<j<k} V_{ijk} } \ .
\end{align}
VMC calculations find an upper bound $E_V$ to an eigenenergy $E_0$ of the 
Hamiltonian by evaluating the expectation value of 
$H$ in a trial wave function, $\Psi_V$:
\begin{align}
E_V = \frac{\langle \Psi_V | H | \Psi_V \rangle}
{\langle \Psi_V | \Psi_V \rangle} \geq E_0  \ .
\end{align}
Parameters in $\Psi_V$ are varied to minimize $E_V$, and the lowest value 
is taken as the approximate energy.
The multidimensional integral is evaluated using standard Metropolis
MC techniques~\cite{MR2T2}, hence the VMC designation.
A good trial function is given by~\cite{WPCP00}
\begin{align}
   |\Psi_V\rangle = {\cal S}\prod_{i<j}^A
      \left[1 + U_{ij} + \sum_{k\neq i,j}^{A}\tilde{U}_{ijk} \right]
                    |\Psi_J\rangle \ ,
\label{eq:psiv}
\end{align}
where $U_{ij}$ and $\tilde{U}_{ijk}$ are non-commuting two- and three-body 
correlation operators induced by the dominant parts of $v_{ij}$ and $V_{ijk}$, 
respectively; ${\cal S}$ is a symmetrizer, and the Jastrow wave function 
$\Psi_J$ is
\begin{align}
   |\Psi_J\rangle = \prod_{i<j}f_c(r_{ij}) |\Phi_A(J^\pi;T T_z)\rangle \ .
\end{align}
Here the single-particle $A$-body wave function $\Phi_A(J^\pi;T T_z)$ is fully 
antisymmetric and has the total spin, parity, and isospin quantum numbers 
of the state of interest, while the product over all pairs of the central 
two-body correlation $f_c(r_{ij})$ keeps nucleons apart to avoid the strong
short-range repulsion of the interaction.
The long-range behavior of $f_c$ and any single-particle radial dependence
in $\Phi_A$ (which, to ensure translational invariance, is written using 
coordinates relative to the center of mass of the $s$-shell core)
control the finite extent of the nucleus.
For $p$-shell nuclei, there are actually three different central pair
correlation functions $f_c$: $f_{ss}$, $f_{sp}$, and $f_{pp}$,
depending on whether both particles are in the $s$-shell core ($ss$), both
in the $p$-shell valence regime ($pp$), or one in each ($sp$).

The two-body correlation operator has the structure 
\begin{align}
   U_{ij} = \sum_{p=2,6} u_{p}(r_{ij}) O^{p}_{ij} \ ,
\end{align}
where the $O^{p}_{ij}$ are the leading spin, isospin, spin-isospin, tensor,
and tensor-isospin operators in $v_{ij}$.
The radial shapes of $f_c(r)$ and $u_p(r)$ are obtained by numerically 
solving a set of six Schr\"{o}dinger-like equations: 
two single-channel for $S=0$, $T=0$ or 1,
and two coupled-channel for $S=1$, $T=0$ or 1, with the latter producing
the important tensor correlations~\cite{W91}.
These equations contain the bare $v_{ij}$ and parametrized Lagrange 
multipliers to impose long-range boundary conditions of exponential decay
and tensor/central ratios.

Perturbation theory is used to motivate the three-body correlation operator
\begin{align}
\tilde{U}_{ijk} = -\epsilon \tilde{V}_{ijk}(\tilde{r}_{ij},
                     \tilde{r}_{jk}, \tilde{r}_{ki}) \ ,
\end{align}
where $\tilde{r}=yr$, $y$ is a scaling parameter, $\epsilon$ is a small strength parameter, and $\tilde{V}_{ijk}$ includes the anticommutator part of two-pion exchange and the phenomenological short-range repulsion in the three-nucleon potential.
Consequently, $\tilde{U}_{ijk}$ has the same spin, isospin, and tensor dependence that $\tilde{V}_{ijk}$ contains.

The variational parameters in $f_{ss}$, $U_{ij}$, and $\tilde{U}_{ijk}$ have 
been chosen to minimize the energy of the $s$-shell nucleus $^4$He.
For the $p$-shell nuclei $^6$He and $^{12}$C, these parameters are kept
relatively constant while the additional parameters that enter $f_{sp}$,
$f_{pp}$, and the single-particle radial behavior of $\Phi_A$ have been 
adjusted to minimize the energy of these systems subject to the constraint
that the proton and neutron rms radii are close to those obtained from more
sophisticated Green's function MC calculations~\cite{PPCPW97,WPCP00}.

The wave function samples used here are generated by following a
random walk guided by  $\Psi_V$ for each nucleus.
After an initial randomization, a move is attempted, where each particle
is randomly shifted within a box of 1.0-1.4~fm in size;  $\Psi_V$ is
evaluated and compared to the previous configuration, with the move 
being accepted or rejected according to the Metropolis algorithm.
After ten attempted moves, the configuration is saved, including the 
$x,y,z$ coordinates of each particle (the center of mass is set to zero)
and the probability for each particle to be either a neutron or a proton.
The size of the box gives an acceptance rate of $\sim$50\% and we have generated 500\,000 configurations for $^4$He, 200\,000 for $^6$He, and 40\,000 for $^{12}$C.
In the calculations below, we generally use 40\,000 configurations for each nucleus.

In order to estimate the contributions of the three-nucleon
  interactions, we show the breakdown of the expectation values
  of the Hamiltonian for the ground state of $^{12}$C. The total energy 
  is $\left<H\right>=-64.4(0.3)$~MeV; $\left<T\right>=404.4(2.3)$~MeV for the kinetic energy,
  $\left<V_2\right>=-465.4(2.1)$~MeV for the two-nucleon potential term,
  $\left<V_C\right>=8.1(0.1)$~MeV for the Coulomb potential term,
  and $\left<V_3\right>=-11.5(0.2)$ MeV for the three-nucleon potential terms
  \footnote{We note that the VMC wave function used here is the starting point for a Green's function Monte Carlo (GFMC) calculation that produces a ground state energy of $-93.3(0.4)$~MeV for $^{12}$C~\cite{Carlson15}.}.
  The three-body force is regarded as essential for a detailed reproduction
  of the nuclear binding energy as $\left<V_3\right>$
  is about 20\% of $\left<H\right>$, while it can be assumed to be negligible
  in the description of reaction processes as
  $\left<V_3\right>$ is about 2\% of $\left<V_2\right>$.
  We also note that the contribution of the three-nucleon force is much smaller
  than that of the two-nucleon force.
  In the ground state of $^{12}$C, the contribution from the two-nucleon interaction is about $-7 (\approx \left<V_2\right>/66)$~MeV/pair, whereas that from the three-nucleon force is $-0.05 (\approx \left<V_3\right>/220)$~MeV/triple.
  For the three-nucleon force to become effective in a collision between
  the projectile and target nuclei, three nucleons must come very close to each other. However, taking into account the strong short-range repulsion in the two-nucleon interaction, such close configurations of three-nucleons are strongly suppressed. Although there are indeed many possible three-nucleon combinations spanning the projectile and target nuclei, their interaction is intrinsically weak and further blocked by the short-range repulsion. Therefore, they are unlikely to provide a significant contribution, except possibly at large scattering angles.

Table~\ref{rms.tab} lists the rms radii of
the wave functions employed in this paper.
The rms radii for protons are consistent with the point-proton radii
deduced from the charge radius measurements~\cite{Angeli13}.
The neutron rms radii $^{6}$He is spatially extended, exhibiting
neutron halo structure.
Appendix~\ref{VMC.density} displays the one- and two-body VMC densities of $^{4}$He, $^{6}$He, and $^{12}$C. 
More details can be found in Ref.~\cite{Piarulli23}.

\begin{table}[hbt]
  \centering
  \caption{Root-mean-square matter, neutron, and proton radii, in units of~fm,  of the VMC wave functions.   40\,000 MC samples are used. The values in parentheses denote uncertainties of the MC
    integration.  The experimental point-proton radii are taken from Ref.~\cite{Angeli13}.}
  \begin{tabular}{cccccc}
\hline\hline    
Nucleus && $r_m$ & $r_n$ &$r_p$& $r_p$ (Expt.) \\ 
\hline
$^{12}$C&& 2.35(1)& 2.35(1) &2.35(1)& 2.326$\pm$0.002\\
$^{4}$He&& 1.44(1)&1.44(1) &1.44(1) & 1.455$\pm$0.003\\
$^{6}$He&& 2.54(1)& 2.80(1)& 1.93(1)& 1.92$\pm$0.01\\
\hline\hline
  \end{tabular}
\label{rms.tab}
\end{table}

\section{Monte Carlo integration of phase-shift function}
\label{numerical.sec}

Let $\rho^P(\rho^T)$ be the $A_P(A_T)$-body density
of the projectile (target) nucleus:
\begin{align}
  &\rho^P(\{\bm{r}\})=\left<\Psi^P_0\right|
  \prod_{i=1}^{A_P}\delta(\bm{r}_i-\bm{r}_i^{P})
  \left|\Psi^P_0\right>,\\
  &\rho^T(\{\bm{r}^{\prime}\})=\left<\Psi^T_0\right|
  \prod_{i=1}^{A_T}\delta(\bm{r}_i^{\prime}-\bm{r}_i^{T})
  \left|\Psi^T_0\right>.
  \label{a-body.eq}
\end{align}
Here, $\{\bm{r}\}=\{\bm{r}_1,\dots,\bm{r}_{A_P}\}$ and $\{\bm{r}^{\prime}\}=\{\bm{r}_1^{\prime},\dots,\bm{r}_{A_T}^{\prime}\}$.  $\{\bm{s}\}$ and $\{\bm{s}^{\prime}\}$ are also defined in  the same way as $\{\bm{s}^P\}$ and $\{\bm{s}^{T}\}$.  The psf of Eq.~(\ref{psf.final}) is obtained by 
the following multiple integration 
\begin{align}
  e^{i\chi(\bm{b})} &=\idotsint d\bm{r}_1\dots d\bm{r}_{A_P}
                                        d\bm{r}_1^{\prime}\dots d\bm{r}_{A_T}^{\prime}\notag \\      & \times \rho^P(\{\bm{r}\})\rho^T(\{\bm{r}^{\prime}\})e^{i\chi_N^{\rm tot}(\bm{b},\{\bm{s}\},\{\bm{s}^{\prime} \})} e^{i\Delta\chi_C(\bm{b},\{\bm{s}\},\{\bm{s}^{\prime} \})}. \notag \\
\label{def.psf}
\end{align}

This $3(A_P+A_T)$-dimensional integration is in general possible only with the help of  MCI as performed in Refs.~\cite{Varga02,Kaki12,Nagahisa18}. A general form of the integration is 
\begin{align}
  \int p(x)g(x) dx,
\end{align}
where $x$ stands for a set of multiple integration variables and a weight function $p(x)$ is non-negative satisfying $\int p(x)dx=1$.  For a sufficiently large $N$ we obtain 
\begin{align}
\int p(x)g(x) dx \approx \frac{1}{N}\sum_{i=1}^N g(\bar{x}_i),
\end{align}
 provided that  the MC configurations, $\bar{x}_1,\bar{x}_2,\ldots, \bar{x}_N$,
are generated by the Metropolis sampling method according to the distribution $p(x)$. 
The convergence of the integral is in general slow, so that one has to take a large value for the number of the configurations $N$. In what follows $N$ is chosen to be 
40\,000  for all the cases of $^4$He, $^6$He, and $^{12}$C.

\section{Results}
\label{results.sec}

\subsection{Medium- to high-energy nuclear collisions}
\label{kinematics.sec}

\begin{figure}[ht]
\begin{center}
  \epsfig{file=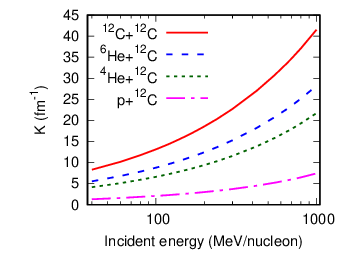, scale=1.4}
  \caption{Wave numbers
of $^{12}$C$+^{12}$C, $^{6,4}$He+$^{12}$C and $p+^{12}$C scatterings
as a function of the incident energy per nucleon.}
    \label{relK.fig}
  \end{center}
\end{figure}

Now we consider a collision of
a projectile nucleus  on a  target nucleus at rest.
The incident energy per nucleon, $E$,  is  $40 \leq E \leq 1000$~MeV,
which corresponds to about 0.3 to 0.9 times  the speed of light in vacuum according to
\begin{align}
\frac{v}{c}=\sqrt{1-\left(\frac{m_N c^2}{m_Nc^2+E}\right)^2}.
\end{align} 
This $v$ is identified with $v$ in Eq.~(\ref{eikeq.eq}). The corresponding wave number $K$ is 
\begin{align}
  K&=\frac{A_PA_Tm_Nc^2}{\hbar c}\sqrt\frac{E^2+2Em_Nc^2}{(A_P+A_T)^2m_N^2c^4+2EA_PA_Tm_Nc^2}.
  \label{relK.eq}
\end{align}
See Appendix~\ref{deriv.K} for the derivation of $K$.

Figure~\ref{relK.fig} plots $K$ for $^{12}$C$+^{12}$C, 
  $^{6,4}$He+$^{12}$C, and  $p+^{12}$C scatterings 
as a function of $E$. The figure can be used to check if the eikonal 
approximation, $Ka \gg 1$, is satisfied. 
Here, $a$ is the interaction range of the colliding nuclei; 
it is roughly a sum of the nuclear
radii of the projectile and target nuclei or approximately 3~fm for 
$p$+$^{12}$C scattering.   
For the proton incident energy  $E\gtrsim 200$~MeV, the wave number is $K\gtrsim 3$~fm$^{-1}$, giving $Ka \approx 9 \gg 1$, and  the eikonal condition is  satisfied.  
 For  $^{12}$C$+^{12}$C case,  $K=8.3$~fm$^{-1}$ even at $E=40$~MeV
and  the eikonal approximation is satisfied. 
It should also be noted that the eikonal approximation breaks down
if the low-momentum transfer assumption $q/K \ll 1$ is not satisfied.

\subsection{Coulomb breakup effects}
\label{cbu.sec}

\begin{figure}[ht]
\begin{center}
  \epsfig{file=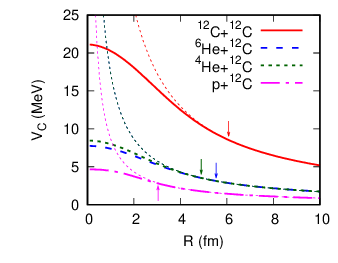, scale=1.4}
  \caption{Coulomb potentials between $^{12}$C$+^{12}$C, $^{6,4}$He+$^{12}$C and $p+^{12}$C 
    as a function of the separation distance. 
    Thin dashed curves are the corresponding point-Coulomb potentials, and arrows indicate the cut-off impact parameters, $b_C$. } 
    \label{Coulpot.fig}
  \end{center}
\end{figure}

The CBU term  $e^{i\Delta\chi_C(\bm{b},\{\bm{s}^P\},\{\bm{s}^T \})}$ of Eq.~(\ref{coulomb.part}) 
depends on  the cut-off radius $b_C$. It is determined by the location where 
 the Coulomb potential between the projectile and target nuclei deviates from the point-Coulomb potential. The argument made there can be confirmed by evaluating the Coulomb potential between them as a function of their separation distance $\bm{R}$.  The 
Coulomb potential can be evaluated with MCI by
\begin{align}
  V_C(\bm{R})&=\left<
  \sum_{i=1}^{A_P}\sum_{j=1}^{A_T}\frac{\epsilon_i\epsilon_je^2}{|\bm{R}+\bm{r}_i^P-\bm{r}_j^T|}
  \right>\notag \\
  &\approx
\frac{1}{N_PN_T}\sum_{k=1}^{N_P}\sum_{l=1}^{N_T}
\sum_{i=1}^{A_P}\sum_{j=1}^{A_T}\frac{\epsilon_i\epsilon_je^2}
    {|\bm{R}+\bar{\bm{r}}_{i,k}^P-\bar{\bm{r}}_{j,l}^T|},
\label{Coulpot.eq}
\end{align}
where $\bar{\bm{r}}_{i,k}^P,\, \bar{\bm{r}}_{j,l}^T$ are appropriately chosen  MC configurations. 
Figure~\ref{Coulpot.fig} compares the computed Coulomb potential  with  the corresponding point-Coulomb potential. 
As expected, the calculated Coulomb potential weakens in the region where the projectile
and target nuclei overlap each other
and coincides with the point-charge Coulomb potential
beyond $b_C$. More precisely, the value of $b_C$  is 
6.07 ($^{12}$C+$^{12}$C),
5.53 ($^{6}$He+$^{12}$C),  
4.89 ($^{4}$He+$^{12}$C), and
3.03 ($p$+$^{12}$C)  in~fm, respectively.
The prescription of the CBU term appears physically quite reasonable. 

\begin{figure}[t]
\begin{center}
  \epsfig{file=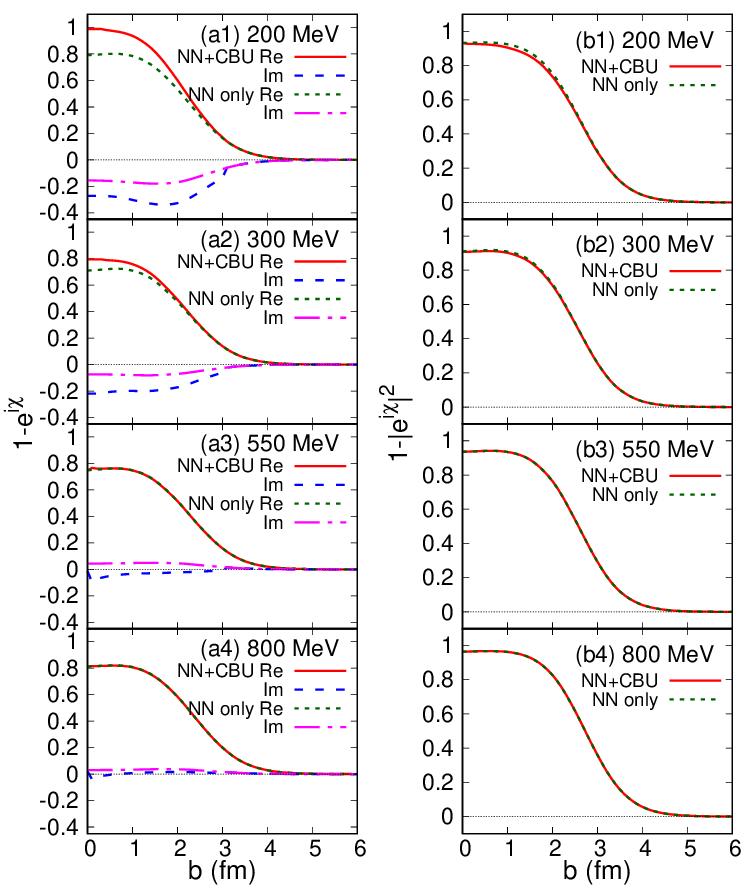, scale=0.68}
  \caption{Impact parameter dependence of  (a) $1-e^{i\chi(b)}$ and (b) $1-|e^{i\chi(b)}|^2$ 
    for $p+^{12}$C scatterings at 
     $E$ =  200,  300,  550, and  800~MeV.}
  \label{psf-p-CBU.fig}
  \end{center}
\end{figure}

\begin{figure}[ht]
\begin{center}
  \epsfig{file=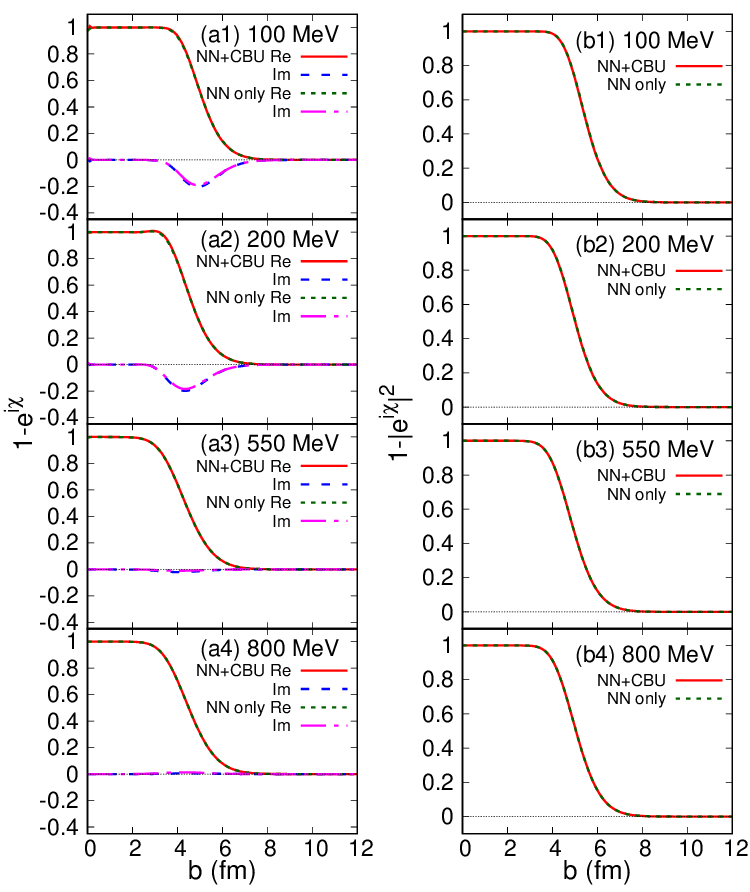, scale=0.68}
  \caption{Same as Fig.~\ref{psf-p-CBU.fig} but for
    $^{12}$C$+^{12}$C scatterings at incident energies of
    100,  200,  550, and  800~MeV/nucleon.  }
    \label{psf-C-CBU.fig}
  \end{center}
\end{figure}

As shown in Eqs. (\ref{elastic.X}) and (\ref{reaction.X}), the $b$-dependence of $1-e^{i\chi(\bm{b})}$ 
and  $1-|e^{i\chi(\bm{b})}|^2$ respectively determines the elastic differential cross section and the total reaction cross section. 
In general the psf   depends on $\bm{b}$. However,   it turns out to be a function of $b$ 
in the present case involving nuclei with zero total angular momentum.
Figure~\ref{psf-p-CBU.fig} displays the $b$-dependence of $1-e^{i\chi(b)}$ (left) and  $1-|e^{i\chi(b)}|^2$ (right) for  $p+^{12}$C scatterings. 
The amplitude of $1-e^{i\chi(\bm{b})}$ is enhanced in the internal region at lower incident energies (200 and 300~MeV)
when the CBU contribution is taken into account. 
Though the Coulomb potential is repulsive,
subtracting  the point-Coulomb potential as in Eq.~(\ref{subC.eq}) gives
the attractive effect in total. The CBU effect becomes smaller
as the incident energy increases, that is, $\eta$ decreases.
While there are some differences in the amplitudes,
the CBU effect on the reaction probability is found to be small.

Figure~\ref{psf-C-CBU.fig} is the same as Fig.~\ref{psf-p-CBU.fig}
but for  $^{12}$C+$^{12}$C scatterings.
The CBU effect  is negligible for both the amplitudes and
reaction probabilities.
Since the nucleus-nucleus collision is strongly absorptive in
the region where the two colliding nuclei overlap,
the CBU effect can contribute only around the touching distance
of the two nuclei where the Coulomb potential between them is close to 
the  point-Coulomb potential.

\subsection{Elastic scattering and total reaction cross sections}
\label{elastic-reaction.sec}

\subsubsection{$p+^{12}$C scattering}
\label{resultsC-p.sec}

\begin{figure*}[ht]
\begin{center}
  \epsfig{file=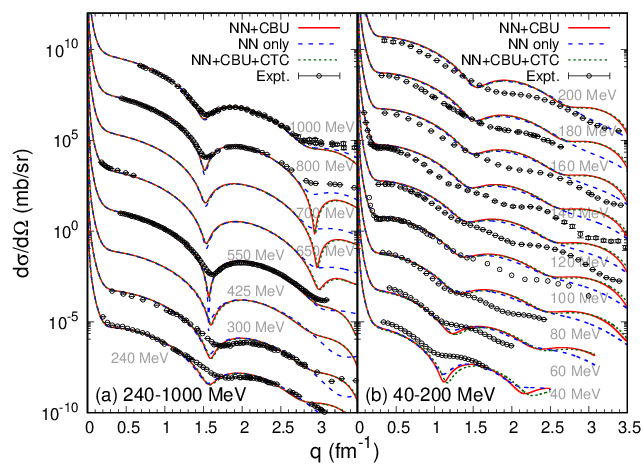, scale=1.3}    
  \caption{ $p+^{12}{\rm C}$ elastic differential  cross sections  at 
    (a) 240, 300, 425, 500, 650, 700, 800, 1000~MeV and
    (b) 40, 60, 80, 100, 120, 140, 160, 180, 200~MeV as a function of
    the momentum transfers $q$. 
    For ease of viewing a factor of 10$^2 (10^{-2})$ is multiplied by the cross section at (a) 650~MeV  or  
    at (b) 120~MeV  with the increase (decrease)  of the incident energy.
    The results with  CBU and CTC  are also shown.    
    The  data are taken from Refs.~\cite{Ieiri87} (40, 60, and 83~MeV),
    \cite{Strauch56} (96~MeV)
    \cite{Meyer83} (122 and 160~MeV)
    \cite{Taylor61,Steinberg65,Emmerson66} (142--145~MeV),
    \cite{Johansson61} (183~MeV),
    \cite{Meyer81} (200~MeV),
    \cite{Meyer88} (250~MeV),
    \cite{Meyer85,Okamoto10} (300~MeV),
    \cite{Hoffmann90} (500~MeV),
    \cite{Azhgirey63} (660~MeV),   
    \cite{Blanpied81} (800~MeV),
    \cite{Palevsky67,Alkhazov72} (1000~MeV).
    Note that  the eikonal approximation
    is fulfilled at $E\gtrsim 200$~MeV. See text for details.}
    \label{dcs-C-p.fig}
  \end{center}
\end{figure*}

\begin{figure}[ht]
\begin{center}
  \epsfig{file=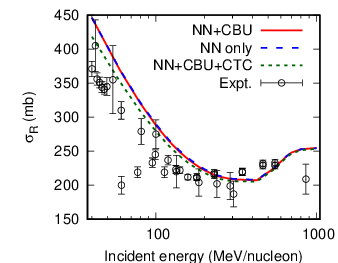, scale=1.3}
  \caption{Energy dependence of the total reaction cross sections
    of $p+^{12}$C collision.
    The cross sections with  CBU and CTC  
    are also shown. The  data are taken from Refs.~\cite{Carlson96,Auce05}.}
    \label{RCS-p.fig} 
  \end{center}
\end{figure}

There are numerous 
data on both elastic scattering and total reaction cross section for the $p+^{12}$C system. 
Figure~\ref{dcs-C-p.fig} displays the $p+^{12}$C elastic differential 
 cross sections
at  $40\leq E \leq 200$~MeV (right) and 
 $240 \leq  E \leq 1000$~MeV (left) incident energies 
as a function of $q$. 
The experimental cross sections at the low incident energies are well reproduced up to the first dip.
As long as the eikonal condition 
is satisfied at $E > 200$~MeV, the theory reproduces 
the experiment  fairly well up to the second dip. 
It should also be noted that the eikonal approximation
is valid for  $q/K \ll 1$ or at least $q\lesssim K/2$.
For example,  $K$ is less than 2~fm$^{-1}$ at $E=100$~MeV (see Fig.~\ref{relK.fig}), so that the agreement between  theory and experiment is limited to $q \lesssim 1$~fm$^{-1}$. 
The CBU contributions are small in the $p+^{12}$C scattering.
We see some differences in the backward angles beyond the second dip,
where the validity of the eikonal approximation is questionable.
The  CTC effect  is also negligible for the $p+^{12}$C elastic scattering.

Figure~\ref{RCS-p.fig} compares the $p+^{12}$C total 
reaction cross sections between theory and experiment.
Both the eikonal and adiabatic 
approximations are fulfilled at $E \gtrsim 200$~MeV.
Since the experimental data are considerably scattered, 
it is only possible to conclude that the theory gives reasonable agreement to the data. 
The CBU contribution is found to be negligible.
In contrast to the elastic differential  cross sections,
the CTC effects are visible at low incident energies, $E<200$~MeV. 
The total reaction cross section reduces by about 20~mb
at $E=50$~MeV, while its reduction is only a few~mb at $E=1000$~MeV.

Looking at both Figs.~\ref{dcs-C-p.fig} and ~\ref{RCS-p.fig} we point out 
the following contradiction. 
The theory very well reproduces the elastic differential cross sections at the forward angles
for $E>200$~MeV, whereas it appears to underestimate the total reaction cross sections by 15--20~mb 
at $E= 300$--600~MeV. The theory  reproduces the elastic differential  cross sections at $E=800$ and 1000~MeV,  but the calculated total reaction cross section is significantly larger than the measured total reaction cross section at about $E=900$~MeV.
Measuring high-quality total reaction cross sections  
is desired before we look for possible reasons in the theoretical calculation.

\subsubsection{$^{12}$C$+^{12}$C scattering}
\label{resultsC-C.sec}

\begin{figure}[ht]
\begin{center}
  \epsfig{file=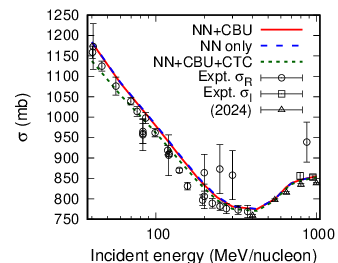, scale=1.3}
  \caption{Energy dependence of the total reaction cross sections
    of $^{12}$C$+^{12}$C collision.
The cross sections with  CBU
    and CTC   are also shown.    
    The data are taken from Refs.~\cite{Takechi05,Perrin82,Zhang02a,fang00,Kox87,Zheng02b,Hostachy88,Jaros78}.}
    \label{RCS-C.fig}
  \end{center}
\end{figure}

\begin{figure}[hbt]
\begin{center}
  \epsfig{file=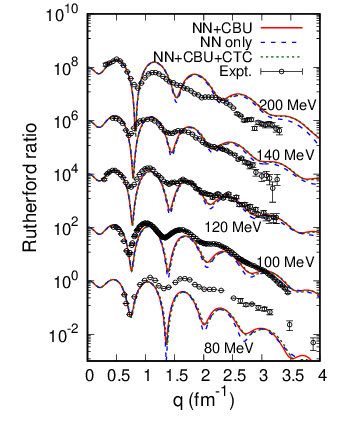, scale=1.4}  
  \caption{Rutherford ratios of  $^{12}{\rm C}+ ^{12}$C elastic differential 
    cross sections at 80, 100, 120, 140, and 200~MeV/nucleon as a function of 
    the momentum transfer $q$. 
    The results with  CBU and CTC are also shown.    
   For ease of viewing a factor of 10$^2$ is multiplied by the cross section at 80~MeV/nucleon   
    with the increase  of the incident energy.
    The data at 86, 100, 120, 135, and 200~MeV/nucleon taken from
    Refs.~\cite{Buenerd81, Hostachy87, Ichihara94, Qu15} are drawn.}
    \label{dcs-C-C.fig}
  \end{center}
\end{figure}

Figure~\ref{RCS-C.fig} displays the  total reaction
cross sections of $^{12}$C+$^{12}$C collisions  calculated at $E$ = 40--1000~MeV.
They are compared to the experimental data
that contain both total reaction and interaction cross sections.  
The latter cross sections are estimated to be smaller
by dozens of mb than the  total reaction 
cross sections~\cite{Kohama08,Takechi14}. 
The agreement between  theory and experiment is attained and in fact 
is much better than  the $p$+$^{12}$C case. Note that the eikonal condition $Ka\gg 1$ is fulfilled 
in the range of $40 \leq E \leq 1000$~MeV
and the adiabatic approximation is safely met for  
the incident energy larger than 80~MeV/nucleon.  
The CBU effect is found to be small.  
The  CTC effect  gives a 
  reduction by about 20~mb at $E=80$~MeV, but it  
  is negligible at high incident energies, giving only a few~mb
  at $E=1000$~MeV.
    The theory reproduces the interaction 
cross section data at high incident energies at $E=800$--900~MeV. 
Very recently, the interaction cross sections  have been measured
above $E=400$~MeV with only small error bars~\cite{Ponnath24}.
The calculated reaction cross sections are slightly larger than 
    experiment, which appears to be very reasonable  considering 
    $\sigma_R\gtrapprox\sigma_I$ at high incident energies.
    This excellent agreement strongly confirms   
the validity of  the present {\it ab initio} Glauber-theory calculation using the realistic VMC wave functions. 
The data of Ref.~\cite{Kox87} largely deviate from both the present results and those of Ref.~\cite{Ponnath24}.

Figure~\ref{dcs-C-C.fig} shows
the  $^{12}{\rm C}+^{12}$C elastic differential  cross sections in  ratio of the Rutherford cross sections at 80--200~MeV/nucleon.
Agreement between theory and experiment is quite satisfactory except for the case of 200~MeV/nucleon. As will be shown in the next subsection, 
the $^4$He$+^{12}$C elastic differential cross sections
at 300~MeV/nucleon are very well reproduced. We 
hope that another accurate measurement at 200--300~MeV/nucleon
will be useful to resolve this issue. Both effects of CBU and  
CTC are very small  even at the low incident energies. 
 The cross sections at the backward angles are better reproduced 
 as the incident energy increases: At 80~MeV/nucleon
 only the cross sections at $q\lesssim 1$~fm$^{-1}$ are reproduced,
 where the low momentum transfer condition $q/K \ll 1$
 for the eikonal approximation is safely fulfilled as $K=11.7$~fm$^{-1}$;
 and at 100--140~MeV/nucleon the cross sections
 at $1 \lesssim q\lesssim 2$~fm$^{-1}$ are better reproduced,
 where $13.1 < K < 15.5$~fm$^{-1}$.
 The discrepancy found at 80~MeV/nucleon may be related to the breakdown
 of the adiabatic approximation as noted in Sect.~\ref{Glauber.sec}.
 
\subsubsection{$^{4}$He$+^{12}$C scattering}
\label{results4He-C.sec}

\begin{figure}[ht]
\begin{center}
  \epsfig{file=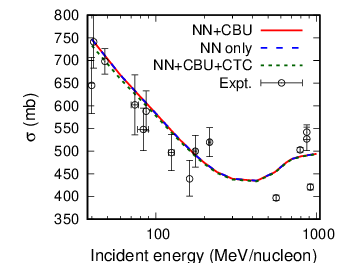, scale=1.2}
  \caption{Same as Fig.~\ref{RCS-C.fig} but for 
    $^{4}{\rm He}+^{12}$C collision.
    The  data are taken from Refs.~\cite{Jaros78, DeVries82, Goekmen84, Tanihata85c, Webber90, Ingemarsson00, Horst17, Horst19}.}
    \label{RCS-4He.fig}
  \end{center}
\end{figure}
  
\begin{figure}[ht]
\begin{center}
  \epsfig{file=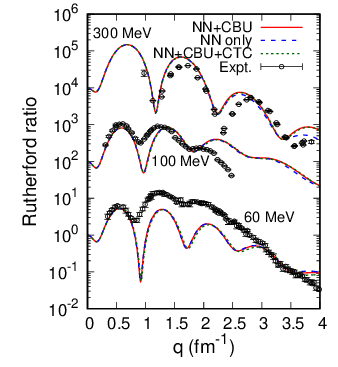, scale=1.2}
  \caption{Same as Fig.~\ref{dcs-C-C.fig} but for $^{4}{\rm He}+^{12}$C
    scatterings at 60, 100, and 300~MeV/nucleon.
    Scaling factors 10$^2$ and 10$^4$ are multiplied by
    the results at 100 and 300~MeV/nucleon, respectively,
    for the sake of visibility. The data are taken from
    Ref.~\cite{John03} for 60~MeV/nucleon,
    Ref.~\cite{Inaba21} for 97~MeV/nucleon,
    and Ref.~\cite{Chaumeaux76} for 342.5~MeV/nucleon.}
    \label{dcs-He4-C.fig}
  \end{center}
\end{figure}

Figure~\ref{RCS-4He.fig} compares the total reaction cross sections of $^{4}$He$+^{12}$C scatterings between theory and experiment. Note that the eikonal condition is fulfilled at all the incident energies. 
The CBU effect is  small. The low energy data at $E\lesssim 100$~MeV
are well reproduced. We cannot say anything about the cross sections
at  $E\gtrsim 550$~MeV because 
the  data are very much  scattered. More accurate data are needed.

Figure~\ref{dcs-He4-C.fig}  compares 
the  $^{4}{\rm He}+^{12}$C elastic differential  cross sections. 
The result  is similar to  the $^{12}$C$+^{12}$C case:
At 60~MeV/nucleon  $K$ is $5.1$~fm$^{-1}$ and the theory can reproduce the data at $q\lesssim 1$~fm$^{-1}$. With the increase of the incident energy to 100~MeV/nucleon 
the data up to larger $q$ values are well reproduced. 
Furthermore the diffraction pattern as well as its magnitude are very well reproduced up to $q \gtrsim 2$~fm$^{-1}$
at  300~MeV/nucleon corresponding to $K=$11.6~fm$^{-1}$.
This agreement is in contrast to the result
of the $^{12}$C$+^{12}$C scattering at 200~MeV/nucleon.  As pointed out in the previous subsection, 
there is  considerable discrepancy between theory and experiment. A careful remeasurement will be needed to clarify the reason for the discrepancy.

\subsubsection{$^{6}$He$+^{12}$C scattering}
\label{results6He-C.sec}

\begin{figure}[ht]
\begin{center}
    \epsfig{file=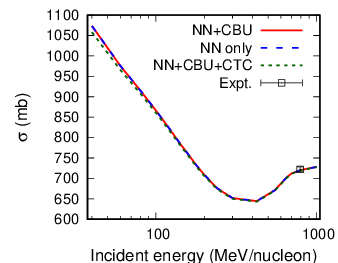, scale=1.2}
    \caption{Same as Fig.~\ref{RCS-C.fig} but for
    $^{6}{\rm He}+^{12}$C collision.
      The experimental interaction cross section datum is taken from Ref.~\cite{Tanihata85b}.}
    \label{RCS-6He.fig}
  \end{center}
\end{figure}

\begin{figure}[ht]
\begin{center}
  \epsfig{file=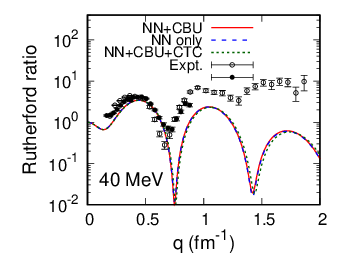, scale=1.2}
  \caption{Same as Fig.~\ref{dcs-C-C.fig} but for
    $^{6}{\rm He}+^{12}$C collision at 40~MeV/nucleon.
    The experimental data are taken from Refs.~\cite{Al-Khalili96c,Lapoux02}}
    \label{dcs-He6-C.fig}
  \end{center}
\end{figure}

Finally we discuss  $^{6}$He+$^{12}$C case  including 
a two-neutron halo nucleus, $^{6}$He.
Figure~\ref{RCS-6He.fig} plots the calculated total reaction cross sections. 
Both CBU and CTC effects can be negligible.  
Only one interaction cross section datum
is available, and the calculated total reaction cross section agrees with 
it very well. 
Since $^{6}$He has no bound excited states, 
the $\sigma_R\approx \sigma_I$ relation holds to a good approximation.  
The Glauber calculation 
is found to  reliably describe high-energy  reactions
involving a spatially extended nuclear system.

Figure~\ref{dcs-He6-C.fig} displays
the $^{6}$He+$^{12}$C elastic differential cross sections  at 40~MeV/nucleon.
Though the incident energy is rather low and $K=4.2$~fm$^{-1}$ at 40~MeV/nucleon,
the calculation reproduces  the experimental data 
up to $q\approx 0.7$~fm$^{-1}$. 
The cross sections at larger $q$ values are not well reproduced similarly 
to the $^{12}$C$+^{12}$C and $^4$He$+^{12}$C cases. This again 
calls for a treatment beyond the adiabatic approximation.
As one would expect from the other systems,
the theory will work better at higher incident energies.

\section{Approximations to phase-shift function}
\label{approx.sec}

\subsection{Cumulant expansion and many-body densities}
\label{cumu.exp}

As already stressed,  the profile function~(\ref{profile.fn}) or the psf of Eq.~(\ref{psf.final}) or (\ref{def.psf}) can in general be calculated only with the MCI.  A systematic way of analyzing 
the psf, known as a cumulant expansion~\cite{Glauber, Hufner81, Ogawa92}, is useful to evaluate some approximate ways to the psf. 
Its leading term, the so-called optical-limit approximation (OLA),
has often been employed because  it requires only one-body densities of the projectile and target nuclei. 
The OLA does not work well, however,  
especially in nucleus-nucleus collisions including 
a halo nucleus~\cite{Bertsch90, Ogawa92, Al-Khalili96a, Al-Khalili96b, Suzuki03}.
Another approximation, the NTG~\cite{NTG,NTG2} 
is also formulated  as a variant of the OLA. Instead, we establish that the cumulant expansion
up to the second order provides  a good approximation to the complete psf.

Since the CBU phase is found to make a minor contribution to the full psf, we neglect it in what follows. The psf $\chi(\bm{b})$ due to the nuclear phase 
is defined by  $i\chi(\bm{b})=\ln G(\bm{b}, 1)$ (see  Eqs.~(\ref{sigma_r.eq}) and (\ref{nucl.part})), where 
\begin{align}
 G(\bm{b}, \lambda)&\equiv\left<  \prod_{i=1}^{A_P}\prod_{j=1}^{A_T}
    \Big(1-\lambda \Gamma_{NN}(\bm{b}_{ij}) \Big)  \right>\notag \\
&=1+\sum_{n=1}^{A_PA_T}\mu_n(\bm{b})\lambda^n
\end{align}  
 is an $A_PA_T$-th  degree polynomial of $\lambda$. $\mu_n(\bm{b})$ is called 
the $n$-th moment and it is expressed in terms of  $1, 2, \cdots, n$-body densities.  For example,  
\begin{widetext}
\begin{align}
\mu_1(\bm{b})&=-\left< \sum_{i=1}^{A_P}\sum_{j=1}^{A_T}\Gamma_{NN}(\bm{b}_{ij})\right>
=-\iint d\bm{r}\,d\bm{r} \rho_1^{P}(\bm{r})\rho_1^T(\bm{r}^\prime)
  \Gamma_{NN}(\bm{b}+\bm{s}-\bm{s}^\prime),\\
\label{mu1ex.eq} 
  \mu_2(\bm{b})&=\left<\sum_{i=1}^{A_P}\sum_{1 \leq j<l}^{A_T}
  \Gamma_{NN}(\bm{b}_{ij})
  \Gamma_{NN}(\bm{b}_{il})  
  \right>
+\left<\sum_{1 \leq i<k}^{A_P}\sum_{j=1}^{A_T}
  \Gamma_{NN}(\bm{b}_{ij})
  \Gamma_{NN}(\bm{b}_{kj})  
  \right>
  +2\left<\sum_{ 1 \le i < k}^{A_P}\sum_ {1 \le j < l}^{A_T}
  \Gamma_{NN}(\bm{b}_{ij})
  \Gamma_{NN}(\bm{b}_{kl})  
  \right>\notag \\
&= \iiint d\bm{r}\,d\bm{r}^\prime\,d\bm{r}^{\prime\prime}\rho_1^P(\bm{r})\rho_2^T(\bm{r}^\prime,\bm{r}^{\prime\prime})
  \Gamma_{NN}(\bm{b}+\bm{s}-\bm{s}^\prime)\Gamma_{NN}(\bm{b}+\bm{s}-\bm{s}^{\prime\prime})\notag\\
  &+\iiint d\bm{r}\,d\bm{r}^\prime\,d\bm{r}^{\prime\prime}\rho_2^P(\bm{r},\bm{r}^\prime)\rho_1^T(\bm{r}^{\prime\prime})
  \Gamma_{NN}(\bm{b}+\bm{s}-\bm{s}^{\prime\prime})
  \Gamma_{NN}(\bm{b}+\bm{s}^\prime-\bm{s}^{\prime\prime})\notag\\
  &+2\iiiint d\bm{r}\,d\bm{r}^\prime\,d\bm{r}^{\prime\prime}\,d\bm{r}^{\prime\prime\prime}\rho_2^P(\bm{r},\bm{r}^\prime)\rho_2^T(\bm{r}^{\prime\prime},\bm{r}^{\prime\prime\prime})
  \Gamma_{NN}(\bm{b}+\bm{s}-\bm{s}^{\prime\prime})\Gamma_{NN}(\bm{b}+\bm{s}^\prime-\bm{s}^{\prime\prime\prime}).
\end{align}
\end{widetext}
Here, $\rho_n$ stands for $n$-body density in which the integration of the spin-isospin coordinates are done as noted before. Figure~\ref{diagram.fig}  illustrates the examples of the relationship between the active profile functions and the relevant densities of the projectile and target nuclei.
A vertex with $n$ dotted lines requires $n$-body density.
In Appendix~\ref{VMC.density} we display the one- and two-body densities of $^{12}$C and $^{4,6}$He  calculated from the VMC wave functions.   It is in general too hard to calculate higher-order moments without the MCI.

\begin{figure}[ht]
\begin{center}
  \epsfig{file=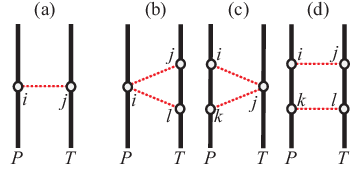, scale=1.2}
  \caption{Schematic illustrations
    for $\mu_1$ (a) and for $\mu_2$ containing three different types
    of collisions (b), (c), and (d). 
    For example, in (a) circles and connecting dotted lines indicate
    the collision of the $i$th nucleon in the projectile ($P$) nucleus
    and the $j$th nucleon in the target ($T$) nucleus.}
    \label{diagram.fig}
  \end{center}
\end{figure}

One is tempted to approximate $e^{i\chi(\bm{b})}$ needed in the scattering amplitude Eq.~(\ref{elastic.X}) with only a few moments, $1+\mu_1(\bm{b})+\cdots$.  
Unfortunately, the series is in general alternating and  $\sum_{n=1}^{A_PA_T} \mu_n(\bm{b})$ very slowly converges~\cite{Ibrahim99}.
Instead, we employ the cumulant expansion defined by
\begin{align}
  \ln G(\bm{b};\lambda)=\sum_{n=1}^\infty\frac{\lambda^n}{n!}\kappa_n(\bm{b}).
\end{align}
The $n$-th cumulant, $\kappa_n(\bm{b})$,  is expressed in terms of the moments, $\mu_k\, (k=1,2,\ldots, n$). For example,  
\begin{align}
  &\kappa_1(\bm{b})=\mu_1(\bm{b}),\notag \\
  &\kappa_2(\bm{b})=2\mu_2(\bm{b})-\mu_1(\bm{b})^2, \notag \\
  &\kappa_3(\bm{b})=6\mu_3(\bm{b})-6\mu_2(\bm{b})\mu_1(\bm{b})+2\mu_1(\bm{b})^3.
  \end{align}
The cumulant expansion gives $e^{i\chi(\bm{b})}$ in terms of the moments as follows:
\begin{widetext}
\begin{align}
 e^{i\chi(\bm{b})}=e^{\ln G(\bm{b};\lambda=1)}
 =\exp\left[\mu_1(\bm{b})+\frac{1}{2}\left(2\mu_2(\bm{b})-\mu_1(\bm{b})^2\right)+\frac{1}{6}\left(6\mu_3(\bm{b})-6\mu_2(\bm{b})\mu_1(\bm{b})+2\mu_1(\bm{b})^3\right)+\dots\right].
\end{align}
\end{widetext}

The moment $\mu_1(\bm{b})$ contains the profile function to the first order.
The psf taking into account $\mu_1(\bm{b})$ only is 
the OLA. Because of its simplicity it has been applied in many cases. 
In fact it works reasonably well for nucleon-nucleus scatterings 
but it is poor  especially for  cases  involving a halo nucleus~\cite{Bertsch90, Ogawa92, Al-Khalili96a, Al-Khalili96b, Suzuki03} and 
the next-order correction  is important. 

To describe the nucleus-nucleus scattering efficiently,
the psf for a nucleus-nucleus scattering
may be described by introducing the nucleon-target
psf, $\chi_{NT}(\bm{b})$,~\cite{NTG,NTG2}:
\begin{align}
  e^{i\chi_{\rm NTG}(\bm{b})}&=
\left<\Psi_0^{P}\right|
  \prod_{i\in P}^{A_P}e^{i\chi_{NT}(\bm{b}+\bm{s}_i^P)}
  \left|\Psi_0^P\right>\notag\\
    &=
  \left<\Psi_0^{P}\right|
  \prod_{i\in P}^{A_P}\left[1-\Gamma_{NT}(\bm{b}+\bm{s}_i^P)\right]
  \left|\Psi_0^P\right>.
  \label{ntgpri.eq}
\end{align}
The nucleon-target profile function $\Gamma_{NT}$ is defined in the same way
as in the $NN$ case  by
\begin{align}
  e^{i\chi_{NT}(\bm{b})}&\equiv 1-\Gamma_{NT}(\bm{b}) \notag \\
  &=\left<\Psi_0^{T}\right|
  \prod_{j\in T}^{A_T}\left[1-\Gamma_{NN}(\bm{b}+\bm{s}_j^T)\right]
  \left|\Psi_0^T\right>.
\end{align}
Since the leading-order approximation of the cumulant expansion works well 
for the nucleon-nucleus scattering, it is  reasonable to assume 
\begin{align}
  1-\Gamma_{NT}(\bm{b})\approx \exp\left[-\int d\bm{r}\, \rho_1^T(\bm{r})\Gamma_{NN}(\bm{b}-\bm{s})\right].
\end{align}
Substituting this expression into Eq.(\ref{ntgpri.eq})
and again taking the leading-order term of the cumulant expansion,
we get the formula 
\begin{widetext}
\begin{align}
e^{i\chi_{\rm NTG}(\bm{b})}=\exp\left\{-\int d\bm{r} \rho_1^P(\bm{r})\left[
1-\exp\left(-\int d\bm{r}^\prime \rho_1^T(\bm{r}^\prime)
  \Gamma_{NN}(\bm{b}+\bm{s}-\bm{s}^\prime)\right)\right]\right\}.
\label{NTG.eq}
\end{align}
\end{widetext}
A symmetrized expression of the above formula  called the NTG approximation is used in this paper 
by taking the geometric mean of Eq.~(\ref{NTG.eq}) 
and the one 
obtained by interchanging $\rho_1^{P}$ and $\rho_1^{T}$. 
The advantage of the NTG  is
that one only needs the same input as the  OLA.
As demonstrated in Refs.~\cite{NTG, NTG2,Horiuchi07, Takechi09, Nagahisa18},
the NTG model gives a better description than the OLA in many cases of
nucleus-nucleus collisions 
and has been used to extract the nuclear radii from
the interaction cross section data~\cite{Kanungo11a,Kanungo11b,Estrade14,Kanungo16,Bagchi19,Tanaka20,Bagchi20,Kaur22}.

At the end of this subsection, we point out that approximating the $A$-body density by a product of the one-body densities and attempting to calculate the psf to all orders  leads practically to  the OLA, that is, nothing better than the OLA. 
This approximation was originally suggested by Glauber~\cite{Glauber} and has very recently been  
revisited~\cite{Shabelski21}. 
Let us assume the $A$-body density to be given by 
\begin{align}
  \rho_{A}(\bm{r}_1,\dots,\bm{r}_A)=
  \prod_{i=1}^{A}\bar{\rho}(\bm{r}_i),
   \label{PDA.eq}
\end{align}
where $\bar{\rho}(\bm{r})=\rho_1(\bm{r})/A$. Then, using Eq.~(\ref{nucl.part})
the psf is given by
\begin{widetext}
\begin{align}
  e^{i\chi(\bm{b})}&=\left(1-\frac{1}{A_PA_T}
  \iint d\bm{r} d\bm{r}^\prime \rho_1^P(\bm{r})\rho_1^T(\bm{r}^\prime)
  \Gamma_{NN}(\bm{b}-\bm{s}+\bm{s}^\prime)\right)^{A_PA_T}\notag \\
  &\approx \exp\left(-\iint  d\bm{r} d\bm{r}^\prime \rho_1^P(\bm{r})\rho_1^T(\bm{r}^\prime)\Gamma_{NN}(\bm{b}-\bm{s}+\bm{s}^\prime)\right),   
\end{align}
\end{widetext}
because $A_PA_T$ is usually significantly large compared to unity. 
The product ansatz (\ref{PDA.eq}) fails to account for higher-order correlations included in the many-body wave function and the resulting psf practically
reduces to the OLA.

\subsection{Evaluation of approximate  phase-shift functions}
\label{eval.sec}

\begin{figure}[ht]
\begin{center}
  \epsfig{file=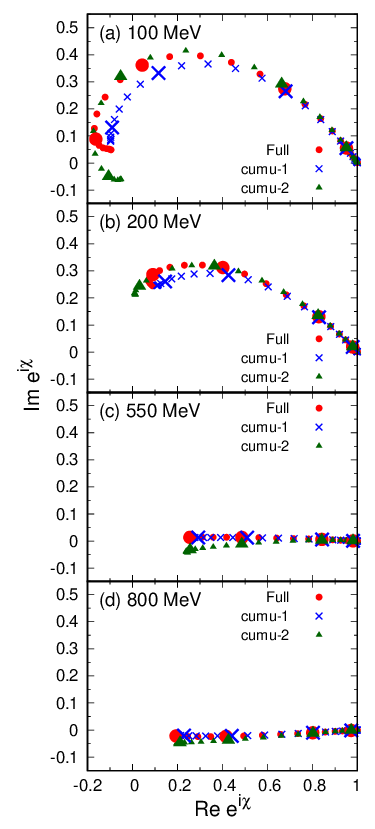, scale=1.2}
  \caption{Phase-shift functions of $p+^{12}$C collision at
    (a) 100, (b) 200, (c) 550, (d) 800~MeV/nucleon on a complex plane.
    The values are plotted by every 0.2~fm and bigger symbols 
    specify the values at $b=1,2,3$ and 4~fm from left to right.}
    \label{psf-p.fig}
  \end{center}
\end{figure}

We compare the approximations discussed in Sect.~\ref{cumu.exp}   by plotting the 
resulting values of  $e^{i\chi(\bm{b})}$ as a function of $b$ on a complex plane.  The psf calculated including up to the $n$-th cumulant is called cumu-$n$ in what follows. Cumu-1 is nothing but the OLA. Full stands for the complete calculation  
including all orders of the cumulants. 
Figure~\ref{psf-p.fig} compares the psf for $p+^{12}$C scatterings.
All the trajectories start at around the origin 
and   move   to unity  with increasing $b$. 
The psf of Full and that of cumu-1 behave differently
at  lower energies, 100 and 200~MeV/nucleon, while the trajectory of cumu-2 
is almost identical to that of Full. 
At 550 and 800~MeV/nucleon, the absolute values of $e^{i\chi(\bm{b})}$ are all 
small and no significant difference appears in the $p+^{12}$C cross sections.

\begin{figure}[ht]
\begin{center}
  \epsfig{file=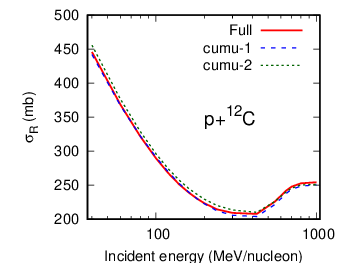,scale=1.2}
  \caption{Comparison of the total reaction cross sections
    obtained by the complete Glauber model calculations
    and various approximate methods for $p+^{12}$C,
    collision as a function of incident energies. See text for details.}
    \label{RCS-p-comp.fig}
  \end{center}
\end{figure}

\begin{figure}[ht]
\begin{center}
    \epsfig{file=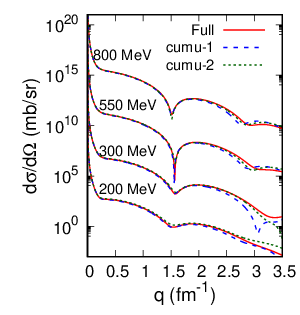, scale=1.5}
  \caption{Comparison of differential elastic-scattering
    cross sections with various approximations for $^{12}{\rm C}+p$ collisions
    as a function of the momentum transfer $q$.
    The cross sections are plotted at 200, 300, 550, and 800~MeV
    from bottom to top.
    For the sake of visibility, some scaling factors are multiplied
    so that the cross sections of adjacent different incident energies
    differ by a factor of 10$^4$.}
    \label{dcs-p-comp.fig}
  \end{center}
\end{figure}

\begin{figure*}[ht]
  \begin{center}
      \epsfig{file=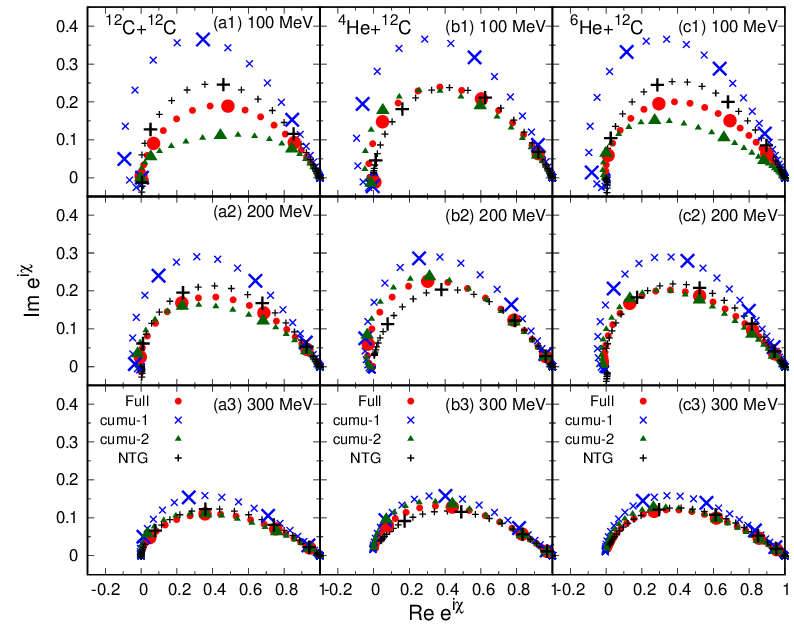, scale=1.2}    
  \caption{Phase-shift functions of (a1)--(a3) $^{12}$C$+^{12}$C, (b1)--(b3) $^{4}$He$+^{12}$C, and (c1)--(c3) $^{6}$He$+^{12}$C collisions in a complex plane.
    The incident energies are chosen
    as (1) 100, (2) 200, and (3) 300~MeV/nucleon.
    The values are plotted by every 0.2~fm and bigger symbols 
    specify the values at $b=3,4,5$ and 6~fm
    for $^{12}$C+$^{12}$C and $^{6}$He+$^{12}$C collisions
    and $b=2,3,4$ and 5~fm for $^{4}$He+$^{12}$C collision
    from left to right.}
    \label{psf.fig}
  \end{center}
\end{figure*}

Figure~\ref{RCS-p-comp.fig} compares the $p+^{12}$C total reaction
cross sections obtained with Full, cumu-1, and cumu-2 calculations.
Both cumu-1 and cumu-2 approximations work well in this case, 
giving  results close to  Full calculation.
Figure~\ref{dcs-p-comp.fig} compares 
 $p+^{12}$C elastic differential cross sections at different incident energies. 
As expected, both cumu-1 and cumu-2 calculations provide us with satisfactory results.

Differences between the approximations become more visible 
in the nucleus-nucleus collisions.
Figure~\ref{psf.fig} compares $e^{i\chi(\bm{b})}$
among different approximations in 
 $^{12}$C$+^{12}$C, $^4$He$+^{12}$C, and $^6$He$+^{12}$C collisions
at the incident energies of 100, 200, and 300~MeV/nucleon.
The trajectories of Full and cumu-1  are quite
different, leading to much difference in the cross sections as shown in the previous section.
The deviation becomes smaller as the incident energy increases.
The psf greatly improves 
when cumu-2 approximation is performed.  
The NTG model also gives better results than cumu-1 
but its psf shows  behavior differently  from cumu-2.

\begin{figure}[ht]
  \begin{center}
  \epsfig{file=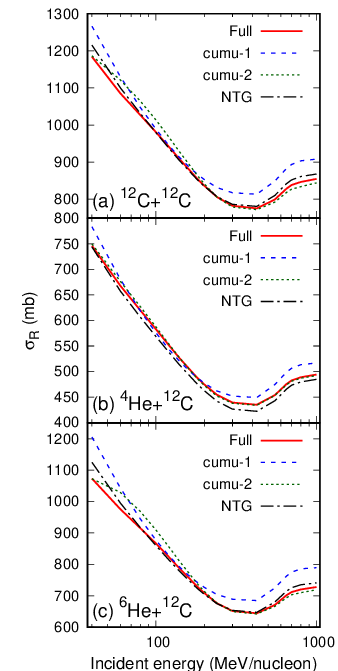,scale=1.2}        
  \caption{Comparison of the total reaction cross sections
    obtained by the complete Glauber model calculations
    and various approximate methods for 
    (a) $^{12}{\rm C}+^{12}$C, (b) $^{4}{\rm He}+^{12}$C, and
    (c) $^{6}{\rm He}+^{12}$C collisions
    as a function of incident energies. See text for details.}
    \label{RCS-comp.fig}
  \end{center}
\end{figure}

Figure~\ref{RCS-comp.fig} compares the total reaction
cross sections of $^{12}{\rm C}+^{12}$C,
$^{4}{\rm He}+^{12}$C, and $^{6}{\rm He}+^{12}$C collisions
obtained with  the psf of  Full, cumu-1, cumu-2, or NTG calculation. 
Deviations from Full Glauber calculations
are found in the total reaction cross sections
of the nucleus-nucleus collisions,
especially, in $^6{\rm He}+^{12}$C and $^{12}{\rm C}+^{12}$C collisions at 
the incident energies $\gtrsim 200$~MeV.
Cumu-2 approximation gives much better results than cumu-1, actually 
almost the same results as Full calculations 
even for  $^{6}$He at the incident energies of $\gtrsim 200$~MeV. 
In general, the NTG model works better than cumu-1. It  
 tends to give larger reaction cross sections  
for  $^{12}$C+$^{12}$C and $^{6}$He+$^{12}$C collisions than Full  results, 
while it always underestimates the $^{4}$He+$^{12}$C reaction cross sections.

Figure~\ref{dcs-C-comp.fig} compares different approximations to predict
$^{12}{\rm C}+^{12}$C and $^{4,6}{\rm He}+^{12}$C elastic differential cross sections.  
Cumu-1 reproduces Full results up to 
the second minimum at $q\approx 1.5$~fm$^{-1}$, and then deviates largely at larger $q$ values.
Cumu-1 results are significantly improved when the second-order cumulant 
 is taken into account. Cumu-2 results show reasonable agreement
with Full calculations up to the backward angles.
We find that the NTG-model  reproduces Full results
up to $q\approx 2$~fm$^{-1}$.
In the case of $^{4}$He$+^{12}$C scattering, it tends to
underestimate the cross sections 
beyond the minimum at $q\approx 1$~fm$^{-1}$.
Cumu-2 always gives the better results than the NTG model, 
showing a good  agreement with Full  results.
However, in the case of $^{6}$He$+^{12}$C, 
Full  cross sections are not reproduced well by Cumu-2 calculations 
at the incident energy of 40~MeV/nucleon. 

\begin{figure*}[ht]
  \begin{center}
  \epsfig{file=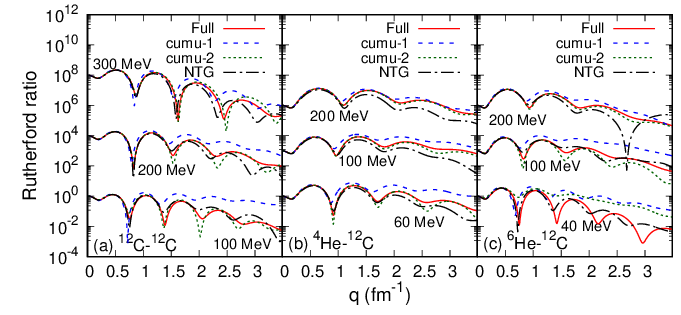, scale=1.4}        
  \caption{Comparison of the Rutherford ratio
    of the differential elastic scattering
    cross sections with various approximations for
    (a) $^{12}{\rm C}+^{12}$C,  $^{4}{\rm He}+^{12}$C, and
    $^{6}{\rm He}+^{12}$C collisions as a function of the
    momentum transfer $q$.
    The incident energies are chosen as
    100, 200, and 300~MeV/nucleon for (a),
    60, 100, and 200~MeV/nucleon for (b), and
    40, 100, and 200~MeV/nucleon for (c).
For the sake of visibility, some scaling factors are multiplied
    so that the cross sections of adjacent different incident energies
    differ by a factor of 10$^4$ for (a) and 10$^3$ for (b) and (c).    
  }
    \label{dcs-C-comp.fig}
  \end{center}
\end{figure*}

\section{Summary and outlook}
\label{conclusion.sec}

We have studied  the elastic differential cross sections and the total reaction cross 
sections of  light projectile nuclei on  $^{12}$C target at intermediate to high incident  
energies within Glauber theory. The projectile nuclei include $p, \, ^4$He, $^6$He, and $^{12}$C. 
The ground-state wave functions of $^{4,6}$He and $^{12}$C were all obtained 
by the variational Monte Carlo method for the realistic Argonne 
$v_{18}$ two-nucleon and Urbana X three-nucleon potentials. 
The  unique advantage of the present approach is that it enables us to
use accurate, sophisticated wave functions of projectile and
target nuclei. 

The nucleus-nucleus elastic scattering
amplitude in Glauber theory is obtained by integrating the profile 
function  over the impact parameter. The profile function 
is the matrix element of the multiple-scattering operator
between the product of the ground-state wave functions of the target
and projectile nuclei. The Coulomb potential contribution to the 
multiple scattering operator has been fully taken into account by separating into 
the point-Coulomb potential term and  the divergence-free term that contributes to the Coulomb breakup.   
The multidimensional integration needed to obtain the profile function has been carried out   
using the Monte Carlo integration method. In this way there has been no need to introduce
any $\it{ad\, hoc}$ approximations or assumptions in the present study once the ground-state 
wave functions of the projectile and target nuclei are given. In this way both the elastic differential cross section and the total reaction cross section  have been  unambiguously evaluated.

The calculated elastic differential cross sections of both $p+^{12}$C and $^{12}$C$+^{12}$C 
 scatterings are in good agreement with experiment provided the incident energy and the scattering angle  are within the eikonal and adiabatic approximations.  
For example, the calculation has very well reproduced the $p+^{12}$C elastic differential cross sections 
up to the second dip at $E \gtrsim 240$~MeV and the $^{12}$C$+^{12}$C elastic differential cross sections 
at $E \gtrsim 100$~MeV/nucleon.. Both CBU and CTC effects are small in the 
elastic differential cross sections. It is worthwhile to stress that the recently measured 
$^{12}$C$+^{12}$C interaction cross sections at 400-1000~MeV/nucleon~\cite{Ponnath24} have been 
very well reproduced by the present calculation. The CTC correction gives a small reduction for the  $^{12}$C+$^{12}$C total reaction cross section at lower energies, e.g., about 20~mb reduction at 
80~MeV/nucleon. As the available experimental  data are fairly scattered, 
accumulating more accurate data is desirable.

Thanks to the full calculation of the phase-shift function we have examined how fast 
that function converges using the cumulant expansion.  For the cases of the projectile-target 
collisions considered here we have found that the full phase-shift function can be well approximated 
up to the second-order in the cumulant expansion. This indicates that both one- and two-body densities 
of the projectile and target nuclei play an important role in the collisions. Therefore the optical-limit 
approximation is  not good enough to describe the nucleus-nucleus collisions. 
For the proton-nucleus case, however, the optical-limit approximation is already fairly good except for the case of the collision including halo nuclei. This encourages us to look into the skin issue of $^{208}$Pb 
using the optical-limit approximation for $p+^{208}$Pb collisions. It looks interesting to extend the previous study~\cite{Horiuchi16} by using the multiple-scattering operator due to the Coulomb interaction as developed here.  

There are numerous data on the interaction cross section since the advance of 
the study on unstable nuclei. It is thus worthwhile establishing a convincing way to calculate  the cross 
section without any {\it ad hoc} assumptions, The  interaction cross section  can be obtained 
in the form $\sigma_I=\int d\bm{b}\, F(\bm{b})$.  In the case of the total reaction cross section, $F(\bm{b})$ ends up 
taking the form,  $\int dx \,p(x)g(\bm{b}, x)$, with the property $p(x)>0, \int dx \,p(x)=1$. As we have shown, 
the Monte Carlo integration has led to the converged 
integral with the importance sampling for a set of $x$. 
 In the case of the interaction cross section, we have to evaluate the type of 
 integral $F(\bm{b})=\int dx \,h(\bm{b},x)$, and the choice of $p(x)$ is not trivial.  If we find  a suitable choice of the importance sampling function $p(x)>0$, the required integral can be approximated by $\frac{1}{N}\sum_{i=1}^N [h(\bm{b},x_i)/p(x_i)].$

In this work, we have neglected the three-nucleon interaction
   between the projectile and target nuclei for simplicity.
   This interaction is expected to play only a minor role, although it may affect the cross sections at large scattering angles. Incorporating three-nucleon interaction terms into the Glauber theory would be an interesting extention of the present study and is left for future work.

\acknowledgments
This work was in part supported by JSPS KAKENHI Grants Nos. 23K22485,
25K07285, 25K01005, and JSPS Bilateral Program No. JPJSBP120247715.
The work of R.~B.~W. is supported by the U.S. Department of Energy, Office of Science, Office of Nuclear Physics, under Contract No. DE-AC02- 06CH11357; VMC calculations were performed on the parallel computers of the Laboratory Computing Resource Center, Argonne National Laboratory and the Argonne Leadership Computing Facility via the INCITE grant ``Ab initio nuclear structure and nuclear reactions.'' Y.~S. is grateful to M. Kimura of RIKEN for his generosity and encouragement.

\appendix

\section{One and two-body densities of VMC wave functions}
\label{VMC.density}

One-body density is defined by
\begin{align}
  \rho_1(\bm{r})&=\left<\Psi_0\right|
  \sum_{i=1}^A  \delta(\bm{r}_i-\bm{r})\left|\Psi_0\right>
\notag \\ 
 & = A\left<\Psi_0\right|
  \delta(\bm{r}_1-\bm{r})\left|\Psi_0\right>.
\label{one-body.density}
\end{align}
Figure~\ref{obd.fig} displays the  one-body densities,
$(4\pi r^2/A)\rho_1(r)$, of $^{12}$C, $^{4}$He, 
and $^6$He of the VMC wave functions. 
As expected, $^{4}$He displays the narrowest distribution peaked at about 1~fm,
while  the distribution of $^{12}$C extends more widely with a peak at about 2~fm. The neutron and proton densities of $^{6}$He are drawn separately. 
Consistently with its $^{4}$He$+n+n$ halo structure,  $^{6}$He exhibits extended 
neutron distribution and its proton density distribution is not as sharp 
as that of $^{4}$He.

\begin{figure}[ht]
\begin{center}
  \epsfig{file=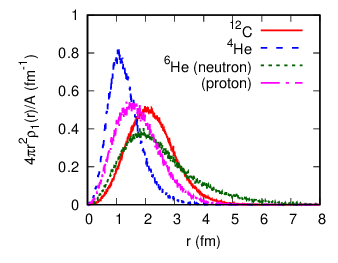, scale=1.4}
  \caption{One-body densities of $^{12}$C,
    $^4$He, and $^{6}$He calculated from the VMC wave functions. The densities of protons and neutrons are identical for  $^{4}$He and $^{12}$C, while they are  separately plotted for $^{6}$He.}
    \label{obd.fig}
  \end{center}
\end{figure}

The two-body density is defined by
\begin{align}
  \rho_2(\bm{r},\bm{r}^\prime)&=\frac{1}{2}\left<\Psi_0\right|
  \sum_{i\neq j}^A\delta(\bm{r}_i-\bm{r})
  \delta(\bm{r}_j-\bm{r}^\prime)\left|\Psi_0\right>\notag \\
 & = \frac{A(A-1)}{2}\left<\Psi_0\right|
 \delta(\bm{r}_1-\bm{r})
  \delta(\bm{r}_2-\bm{r}^\prime)\left|\Psi_0\right>.
\label{two-body.density}
\end{align}
It is a scalar function of $\bm{r}$ and $\bm{r}'$, that is,  $r$, $r^\prime$ and the angle $\theta$ between them because in the present work the spin-parity of $\Psi_0$ is $0^+$. 
Let $A_{12}(\theta)$ be the angular correlation function of two nucleons  defined by  
\begin{align}
  A_{12}(\theta)&=  \frac{2}{A(A-1)}8\pi^2 \sin\theta \notag \\
                   & \ \times 
   \int_0^\infty r^2 dr\int_0^\infty r^{\prime 2} dr^\prime
   \rho_2(r,r^\prime,\theta).
\end{align}
Note that $\int_0^\pi A_{12}(\theta) d\theta=1$. 
Figure~\ref{tbd.fig} displays $A_{12}(\theta)$
for $^{12}$C, $^{4}$He, and $^{6}$He of the VMC wave functions.
The angular correlation function of $^{4}$He has a peak at about $130^\circ$. For the sake of comparison 
the distribution obtained by  the $(0s)^4$ harmonic-oscilator shell-model configuration  is also plotted. 
Note that its  two-body density reads as 
\begin{align}
&  \rho_2(r,r^\prime,\theta)=6\left(\frac{2\beta^2}{\pi^2}\right)^{\frac{3}{2}}
  e^{-\frac{3}{2}\beta\left(r^2+r^{\prime 2}\right)-\beta rr^\prime\cos\theta}.
\end{align}
The $(0s)^4$ angular correlation function with 
the size parameter $\beta=0.52$~fm$^{-2}$ is similar to that of the VMC  wave function. Note, 
however, that the amplitude due to the VMC wave function  reduces at the small  angles due to the short-ranged repulsion of the nuclear force.
In case of $^{6}$He two peaks  appear due to the $p$-shell configuration.
The two peak structure becomes blurred in $^{12}$C
as more $p$-shell nucleons are occupied.
A peak at about $120^\circ$ of $^{12}$C appears consistent with the three-$\alpha$ configuration in its ground state. 

\begin{figure}[ht]
\begin{center}
  \epsfig{file=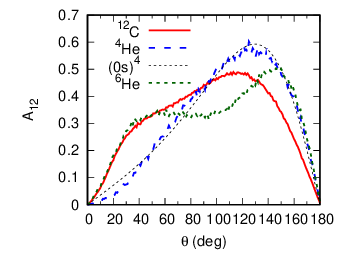, scale=1.4}
  \caption{Angular correlation functions
    of two nucleons  of $^{12}$C
    and $^{4,6}$He calculated from the VMC wave functions. }
    \label{tbd.fig}
  \end{center}
\end{figure}

\section{Derivation of wave number $K$}
\label{deriv.K}

Let a particle with mass $m_1$ (MeV/$c^2$)   impinge on  a particle with  mass $m_2$ (MeV/$c^2$) at rest.   Let the total kinetic energy of the impinging particle be $T_1$ (MeV). The energy-momentum four-vector  of the total system  in the laboratory frame is given  by 
\begin{align}
cp\equiv \left(
\begin{array}{c}
E \\
c\bm p \\
\end{array}
\right)
=\left(
\begin{array}{c}
E_1\\
c{\bm p} \\
\end{array}
\right)
+\left(
\begin{array}{c}
m_2c^2 \\
\bm{0} \\
\end{array}
\right),
\end{align}
while the corresponding four-vector in the center-of-mass system is defined by
\begin{align}
cp'\equiv\left(
\begin{array}{c}
E' \\
\bm{0} \\
\end{array}
\right)
=\left(
\begin{array}{c}
E'_1\\
c\bm{p}' \\
\end{array}
\right)
+\left(
\begin{array}{c}
E'_2 \\
-c\bm{p}' \\
\end{array}
\right).
\end{align}
To relate the momentum $|\bm{p}|$ to $T_1$, we use the following relations:
\begin{align}
&E_1=T_1+m_1c^2,
\label{Eq.B3}\\
&c^2p^2\equiv(E_1+m_2c^2)^2-c^2\bm{p}^2,
\label{Eq.B4}\\
&c^2p'^2\equiv E'^2,\\
&E_1^2-c^2\bm{p}^2=(m_1c^2)^2.
\end{align}
Because of $c^2p^2=c^2p'^2=(m_1c^2+m_2c^2)^2$, we obtain 
\begin{align}
E'^2&=(E_1+m_2c^2)^2-c^2\bm{p}^2\notag \\
&=(E_1+m_2c^2)^2-(E_1^2-(m_1c^2)^2)\notag \\
&=(m_1c^2)^2+(m_2c^2)^2+2E_1m_2c^2.
\label{eq.B4}
\end{align}
Substituting  $E_1$ of Eq.~(\ref{Eq.B3}) to the above equation  we obtain 
\begin{align}
E'=\sqrt{(m_1+m_2)^2c^4+2m_2c^2T_1}.
\end{align}
This leads to 
\begin{align}
c|\bm{p}|=\sqrt{E_1^2-(m_1c^2)^2}=\sqrt{T_1^2+2m_1c^2T_1}.
\end{align}

Assume that the impinging particle moves to the $z$ direction. Under the Lorentz transformation from the center-of-mass frame to the laboratory frame the following equation holds:
\begin{align}
\left(
\begin{array}{c}
E_1+m_2c^2 \\
0\\
0\\
c|\bm{p}|\\
\end{array}
\right)
=
\left(
\begin{array}{cccc}
\gamma & 0 & 0 & \beta \gamma \\
0 & 1 & 0 & 0 \\
0 & 0 & 1 & 0 \\
\beta \gamma & 0 & 0 & \gamma \\
\end{array}
\right)
\left(
\begin{array}{c}
E' \\
0 \\
0 \\
0 \\
\end{array}
\right),
\end{align}
where $\gamma=\frac{1}{\sqrt{1-\beta^2}}$ and $\beta=\frac{v}{c}$ with $v$ being the relative velocity 
between the two frames.  It follows that 
\begin{align}
\beta=\frac{c|\bm{p}|}{E_1+m_2c^2}=\frac{\sqrt{T_1^2+2m_1c^2T_1}}{T_1+(m_1+m_2)c^2}.
\end{align}
The Lorentz transformation for the impinging particle reads 
\begin{align}
\left(
\begin{array}{c}
E'_1 \\
0\\
0\\
-c|\bm{p}'|\\
\end{array}
\right)
=
\left(
\begin{array}{cccc}
\gamma & 0 & 0 & -\beta \gamma \\
0 & 1 & 0 & 0 \\
0 & 0 & 1 & 0 \\
-\beta \gamma & 0 & 0 & \gamma \\
\end{array}
\right)
\left(
\begin{array}{c}
m_2c \\
0 \\
0 \\
0 \\
\end{array}
\right).
\end{align}
The magnitude of the wave number, $K=|\bm{K}|$, in the center-of-mass system is given by
\begin{align}
K&=\frac{|\bm{p}'|}{\hbar}=\frac{\beta\gamma m_2c}{\hbar}\\
&=\frac{m_2c^2}{\hbar c}\sqrt{ \frac{T_1^2+2m_1c^2T_1} {(m_1+m_2)^2c^4+2T_1m_2c^2} }.
\end{align}
Following the definitions in the main text, i.e., by replacing
$m_1\to A_Pm_N, m_2\to A_Tm_N, T_1\to EA_P$,
we get Eq.~(\ref{relK.eq}).

\end{document}